\input harvmac
\def\ev#1{\langle#1\rangle}
\input amssym
\input epsf

\def\unit{\relax{\rm 1\kern-.26em I}}
\def\nada{\relax{\rm 0\kern-.30em l}}



\def\det{{\rm det}}

\noblackbox
\def\IL{\relax{\rm I\kern-.18em L}}
\def\IH{\relax{\rm I\kern-.18em H}}
\def\IR{\relax{\rm I\kern-.18em R}}
\def\IC{\relax\hbox{$\inbar\kern-.3em{\rm C}$}}
\def\IZ{\relax\ifmmode\mathchoice
{\hbox{\cmss Z\kern-.4em Z}}{\hbox{\cmss Z\kern-.4em Z}}
{\lower.9pt\hbox{\cmsss Z\kern-.4em Z}} {\lower1.2pt\hbox{\cmsss
Z\kern-.4em Z}}\else{\cmss Z\kern-.4em Z}\fi}
\def\CM {{\cal M}}

\def\CO {{\cal O}}

\def\CM {{\cal M}}

\def\CO {{\cal O}}

\def\det{{\rm det}}
\def\Tr{{\rm Tr}}

\font\manual=manfnt \def\dbend{\lower3.5pt\hbox{\manual\char127}}

\def\IZ{\relax\ifmmode\mathchoice
{\hbox{\cmss Z\kern-.4em Z}}{\hbox{\cmss Z\kern-.4em Z}}
{\lower.9pt\hbox{\cmsss Z\kern-.4em Z}} {\lower1.2pt\hbox{\cmsss
Z\kern-.4em Z}}\else{\cmss Z\kern-.4em Z}\fi}
\def\half {{1\over 2}}

\def\lfm#1{\medskip\noindent\item{#1}}

\def\Kappa{\kappa_h}

\lref\ISSold{
  K.~A.~Intriligator, N.~Seiberg and S.~H.~Shenker,
  ``Proposal for a simple model of dynamical SUSY breaking,''
  Phys.\ Lett.\ B {\bf 342}, 152 (1995)
  [arXiv:hep-ph/9410203].
}
\lref\MV{
  S.~P.~Martin and M.~T.~Vaughn,
  ``Regularization dependence of running couplings in softly broken
  supersymmetry,''
  Phys.\ Lett.\  B {\bf 318}, 331 (1993)
  [arXiv:hep-ph/9308222].
}
\lref\deGouveaTN{
  A.~de Gouvea, T.~Moroi and H.~Murayama,
  ``Cosmology of supersymmetric models with low-energy gauge mediation,''
  Phys.\ Rev.\  D {\bf 56}, 1281 (1997)
  [arXiv:hep-ph/9701244].
}
\lref\GiveonWP{
  A.~Giveon, A.~Katz and Z.~Komargodski,
  ``On SQCD with massive and massless flavors,''
  arXiv:0804.1805 [hep-th].
}
\lref\DineGM{
  M.~Dine, J.~L.~Feng and E.~Silverstein,
  ``Retrofitting O'Raifeartaigh models with dynamical scales,''
  Phys.\ Rev.\ D {\bf 74}, 095012 (2006)
  [arXiv:hep-th/0608159].
}
\lref\AharonyMY{
 O.~Aharony and N.~Seiberg,
 ``Naturalized and simplified gauge mediation,''
 arXiv:hep-ph/0612308.
}
\lref\BanksMG{
  T.~Banks and V.~Kaplunovsky,
  ``Nosonomy Of An Upside Down Hierarchy Model. 1,''
  Nucl.\ Phys.\  B {\bf 211}, 529 (1983).
}
\lref\ColemanJX{
  S.~R.~Coleman and E.~Weinberg,
  ``Radiative Corrections As The Origin Of Spontaneous Symmetry Breaking,''
  Phys.\ Rev.\  D {\bf 7}, 1888 (1973).
}
\lref\GiudiceNI{
  G.~F.~Giudice and R.~Rattazzi,
  ``Extracting supersymmetry-breaking effects from wave-function
  renormalization,''
  Nucl.\ Phys.\  B {\bf 511}, 25 (1998)
  [arXiv:hep-ph/9706540].
}
\lref\CheungES{
  C.~Cheung, A.~L.~Fitzpatrick and D.~Shih,
  ``(Extra)Ordinary Gauge Mediation,''
  arXiv:0710.3585 [hep-ph].
}
\lref\AGLR{
  N.~Arkani-Hamed, G.~F.~Giudice, M.~A.~Luty and R.~Rattazzi,
  ``Supersymmetry-breaking loops from analytic continuation into  superspace,''
  Phys.\ Rev.\  D {\bf 58}, 115005 (1998)
  [arXiv:hep-ph/9803290].
}
\lref\GiudiceBP{
  G.~F.~Giudice and R.~Rattazzi,
  ``Theories with gauge-mediated supersymmetry breaking,''
  Phys.\ Rept.\  {\bf 322}, 419 (1999)
  [arXiv:hep-ph/9801271].
}
\lref\EssigXK{
  R.~Essig, K.~Sinha and G.~Torroba,
  ``Meta-Stable Dynamical Supersymmetry Breaking Near Points of Enhanced
  Symmetry,''
  JHEP {\bf 0709}, 032 (2007)
  [arXiv:0707.0007 [hep-th]].
}
\lref\WittenNF{
  E.~Witten,
  ``Dynamical Breaking Of Supersymmetry,''
  Nucl.\ Phys.\  B {\bf 188}, 513 (1981).
}
\lref\DineXK{
  M.~Dine, Y.~Nir and Y.~Shirman,
  ``Variations on minimal gauge mediated supersymmetry breaking,''
  Phys.\ Rev.\  D {\bf 55}, 1501 (1997)
  [arXiv:hep-ph/9607397].
}
\lref\ShihAV{
  D.~Shih,
  ``Spontaneous R-symmetry breaking in O'Raifeartaigh models,''
  arXiv:hep-th/0703196.
}
\lref\OR{
  L.~O'Raifeartaigh,
  ``Spontaneous Symmetry Breaking For Chiral Scalar Superfields,''
  Nucl.\ Phys.\  B {\bf 96}, 331 (1975).
}
\lref\IntriligatorCP{
  K.~Intriligator and N.~Seiberg,
  ``Lectures on Supersymmetry Breaking,''
  arXiv:hep-ph/0702069.
}
\lref\OoguriPJ{
  H.~Ooguri and Y.~Ookouchi,
  ``Landscape of supersymmetry breaking vacua in geometrically realized gauge
  theories,''
  Nucl.\ Phys.\  B {\bf 755}, 239 (2006)
  [arXiv:hep-th/0606061].
}
\lref\MeadeWD{
  P.~Meade, N.~Seiberg and D.~Shih,
  ``General Gauge Mediation,''
  arXiv:0801.3278 [hep-ph].
}
\lref\IScomments{
  K.~A.~Intriligator and M.~Sudano,
  ``Comments on General Gauge Mediation,''
  arXiv:0807.3942 [hep-ph].
} 

\lref\IntriligatorNE{
  K.~A.~Intriligator and P.~Pouliot,
  ``Exact superpotentials, quantum vacua and duality in supersymmetric SP(N(c))
  gauge theories,''
  Phys.\ Lett.\  B {\bf 353}, 471 (1995)
  [arXiv:hep-th/9505006].
}
\lref\KaplunovskyYX{
  V.~Kaplunovsky,
  ``Nosonomy Of An Upside Down Hierarchy Model. 2,''
  Nucl.\ Phys.\  B {\bf 233}, 336 (1984).
}
\lref\DineXT{
  M.~Dine and J.~Mason,
  ``Gauge mediation in metastable vacua,''
  arXiv:hep-ph/0611312.
}
\lref\FrancoES{
  S.~Franco and A.~M.~Uranga,
  ``Dynamical SUSY breaking at meta-stable minima from D-branes at obstructed
  geometries,''
  JHEP {\bf 0606}, 031 (2006)
  [arXiv:hep-th/0604136].
}
\lref\IntriligatorIF{
  K.~Intriligator,
  ``IR free or interacting? A proposed diagnostic,''
  Nucl.\ Phys.\  B {\bf 730}, 239 (2005)
  [arXiv:hep-th/0509085].
}
\lref\AmaritiAM{
  A.~Amariti, L.~Girardello and A.~Mariotti,
  ``Meta-stable $A_n$ quiver gauge theories,''
  arXiv:0706.3151 [hep-th].
}
\lref\EGMS{
  E.~Gorbatov and M.~Sudano,
  ``Sparticle Masses in Higgsed Gauge Mediation,''
  arXiv:0802.0555 [hep-ph].
}
\lref\IntriligatorAX{
  K.~A.~Intriligator, R.~G.~Leigh and M.~J.~Strassler,
  ``New examples of duality in chiral and nonchiral supersymmetric gauge
  theories,''
  Nucl.\ Phys.\  B {\bf 456}, 567 (1995)
  [arXiv:hep-th/9506148].
}
\lref\MartinVX{
  S.~P.~Martin,
  ``Two-loop effective potential for a general renormalizable theory and
  softly broken supersymmetry,''
  Phys.\ Rev.\  D {\bf 65}, 116003 (2002)
  [arXiv:hep-ph/0111209].
}

\lref\ForsteZC{
  S.~Forste,
  ``Gauging flavour in meta-stable SUSY breaking models,''
  Phys.\ Lett.\  B {\bf 642}, 142 (2006)
  [arXiv:hep-th/0608036].
}

\lref\IntriligatorNE{
  K.~A.~Intriligator and P.~Pouliot,
  ``Exact superpotentials, quantum vacua and duality in supersymmetric SP(N(c))
  gauge theories,''
  Phys.\ Lett.\  B {\bf 353}, 471 (1995)
  [arXiv:hep-th/9505006].
}
\lref\IntriligatorJJ{
  K.~Intriligator and B.~Wecht,
  ``The exact superconformal R-symmetry maximizes a,''
  Nucl.\ Phys.\  B {\bf 667}, 183 (2003)
  [arXiv:hep-th/0304128].
}
\lref\PouliotSK{
  P.~Pouliot and M.~J.~Strassler,
  ``A Chiral $SU(N)$ Gauge Theory and its Non-Chiral $Spin(8)$ Dual,''
  Phys.\ Lett.\ B {\bf 370}, 76 (1996)
  [arXiv:hep-th/9510228].
}
\lref\AmaritiVK{
  A.~Amariti, L.~Girardello and A.~Mariotti,
  ``Non-supersymmetric meta-stable vacua in SU(N) SQCD with adjoint matter,''
  JHEP {\bf 0612}, 058 (2006)
  [arXiv:hep-th/0608063].
}
\lref\SeibergBZ{
  N.~Seiberg,
  ``Exact results on the space of vacua of four-dimensional SUSY gauge
  theories,''
  Phys.\ Rev.\ D {\bf 49}, 6857 (1994)
  [arXiv:hep-th/9402044].
}
\lref\MartinZB{
  S.~P.~Martin,
  ``Generalized messengers of supersymmetry breaking and the sparticle mass
  spectrum,''
  Phys.\ Rev.\  D {\bf 55}, 3177 (1997)
  [arXiv:hep-ph/9608224].
}
\lref\ISS{
  K.~Intriligator, N.~Seiberg and D.~Shih,
  ``Dynamical SUSY breaking in meta-stable vacua,''
  JHEP {\bf 0604}, 021 (2006)
  [arXiv:hep-th/0602239].
}
\lref\WittenKV{
  E.~Witten,
  ``Mass Hierarchies In Supersymmetric Theories,''
  Phys.\ Lett.\  B {\bf 105}, 267 (1981).
}
\lref\AffleckXZ{
  I.~Affleck, M.~Dine and N.~Seiberg,
  ``Dynamical Supersymmetry Breaking In Four-Dimensions And Its
  Phenomenological Implications,''
  Nucl.\ Phys.\  B {\bf 256}, 557 (1985).
}
\lref\KutasovSS{
  D.~Kutasov, A.~Schwimmer and N.~Seiberg,
  ``Chiral Rings, Singularity Theory and Electric-Magnetic Duality,''
  Nucl.\ Phys.\  B {\bf 459}, 455 (1996)
  [arXiv:hep-th/9510222].
}
\lref\WittenDF{
  E.~Witten,
  ``Constraints On Supersymmetry Breaking,''
  Nucl.\ Phys.\ B {\bf 202}, 253 (1982).
}
\lref\KutasovVE{
  D.~Kutasov,
  ``A Comment on duality in N=1 supersymmetric nonAbelian gauge theories,''
  Phys.\ Lett.\  B {\bf 351}, 230 (1995)
  [arXiv:hep-th/9503086].
}
\lref\NSd{
  N.~Seiberg,
  ``Electric - magnetic duality in supersymmetric nonAbelian gauge theories,''
  Nucl.\ Phys.\ B {\bf 435}, 129 (1995)
  [arXiv:hep-th/9411149].
}
\lref\KutasovNP{
  D.~Kutasov and A.~Schwimmer,
  ``On duality in supersymmetric Yang-Mills theory,''
  Phys.\ Lett.\  B {\bf 354}, 315 (1995)
  [arXiv:hep-th/9505004].
}
\lref\ISSold{
  K.~A.~Intriligator, N.~Seiberg and S.~H.~Shenker,
  ``Proposal for a simple model of dynamical SUSY breaking,''
  Phys.\ Lett.\ B {\bf 342}, 152 (1995)
  [arXiv:hep-ph/9410203].
}
\lref\IntriligatorIF{
  K.~Intriligator,
  ``IR free or interacting? A proposed diagnostic,''
  Nucl.\ Phys.\  B {\bf 730}, 239 (2005)
  [arXiv:hep-th/0509085].
}
\lref\ISSii{
  K.~Intriligator, N.~Seiberg and D.~Shih,
  ``Supersymmetry Breaking, R-Symmetry Breaking and Metastable Vacua,''
  JHEP {\bf 0707}, 017 (2007)
  [arXiv:hep-th/0703281].
}
\lref\BrodieVV{
  J.~H.~Brodie, P.~L.~Cho and K.~A.~Intriligator,
  ``Misleading anomaly matchings?,''
  Phys.\ Lett.\  B {\bf 429}, 319 (1998)
  [arXiv:hep-th/9802092].
}
\lref\RayWK{
  S.~Ray,
  ``Some properties of meta-stable supersymmetry-breaking vacua in Wess-Zumino
  models,''
  Phys.\ Lett.\  B {\bf 642}, 137 (2006)
  [arXiv:hep-th/0607172].
}
\lref\ForsteZC{
  S.~Forste,
  ``Gauging flavour in meta-stable SUSY breaking models,''
  Phys.\ Lett.\  B {\bf 642}, 142 (2006)
  [arXiv:hep-th/0608036].
}
\lref\DineVC{
  M.~Dine, A.~E.~Nelson and Y.~Shirman,
  ``Low-Energy Dynamical Supersymmetry Breaking Simplified,''
  Phys.\ Rev.\  D {\bf 51}, 1362 (1995)
  [arXiv:hep-ph/9408384].
}
\lref\DineAG{
  M.~Dine, A.~E.~Nelson, Y.~Nir and Y.~Shirman,
  ``New tools for low-energy dynamical supersymmetry breaking,''
  Phys.\ Rev.\  D {\bf 53}, 2658 (1996)
  [arXiv:hep-ph/9507378].
}
\lref\DineYW{
  M.~Dine and A.~E.~Nelson,
  ``Dynamical supersymmetry breaking at low-energies,''
  Phys.\ Rev.\  D {\bf 48}, 1277 (1993)
  [arXiv:hep-ph/9303230].
}
\lref\NSd{
  N.~Seiberg,
  ``Electric - magnetic duality in supersymmetric nonAbelian gauge theories,''
  Nucl.\ Phys.\ B {\bf 435}, 129 (1995)
  [arXiv:hep-th/9411149].
}
\lref\CsakiZB{
  C.~Csaki, M.~Schmaltz and W.~Skiba,
  ``Confinement in N = 1 SUSY gauge theories and model building tools,''
  Phys.\ Rev.\  D {\bf 55}, 7840 (1997)
  [arXiv:hep-th/9612207].
}

\lref\GKKS{
  A.~Giveon, A.~Katz, Z.~Komargodski and D.~Shih,
  ``Dynamical SUSY and R-symmetry breaking in SQCD with massive and massless
  flavors,''
  arXiv:0808.2901 [hep-th].
}

\def\gmed{\refs{\DineYW\DineVC\DineAG-\DineXK}}

\newbox\tmpbox\setbox\tmpbox\hbox{\abstractfont }
\Title{\vbox{\baselineskip12pt \hbox{UCSD-PTH-08-06}}}
{\vbox{\centerline{Surveying Pseudomoduli:}
\vskip8pt\centerline{the Good, the Bad and the Incalculable}}}
\smallskip
\centerline{Kenneth Intriligator$^{1}$, 
David Shih$^2$, and Matthew Sudano$^{1}$}
\smallskip
\bigskip
\centerline{$^1${\it Department of Physics, University of
California, San Diego, La Jolla, CA 92093 USA}}
\medskip
\centerline{$^2${\it School of Natural Sciences, Institute for
Advanced Study, Princeton, NJ 08540 USA}}
\bigskip
\vskip 1cm

\noindent 
We classify possible types of pseudomoduli which arise when supersymmetry is dynamically broken in infrared-free low-energy theories. 
We show that, even if the pseudomoduli potential is generated only at higher loops, there is a regime where the potential can be simply determined from a combination of one-loop running data.   In this regime, we compute whether the potential for the various types of pseudomoduli is safe, has a dangerous runaway to the UV cutoff of the low-energy theory, or is incalculable.    Our results are applicable to building new models of supersymmetry breaking.  We apply the results to survey large classes of models. 

\bigskip

\Date{September 2008}

\newsec{Introduction}

\subsec{Motivation}

Dynamical supersymmetry breaking (DSB) is a promising scenario \WittenNF\ for explaining the huge hierarchy between the weak scale and the Planck scale.  Only very special examples seem to exhibit complete DSB at weak electric coupling  (see e.g.\ \AffleckXZ\ and references cited therein).  Another framework for DSB, which leads to many new classes of examples, is via theories with IR-free magnetic duals\foot{The IR phase must be under control, as seen in the original, still inconclusive, example  \ISSold.}, with SUSY  broken at tree-level in the dual.    Accepting long-lived metastable vacua further expands the classes of theories with DSB, among them massive SQCD, which suggests that   metastable DSB can be common, even generic, in field theory and string theory \ISS. See
\IntriligatorCP\ for a recent review and references.     

In analyzing such theories, one must always pay attention to the tree-level flat directions in the potential.  Such ``pseudomoduli'' fields -- which we will collectively denote throughout by $\Phi$ -- are always present in the low-energy F-term SUSY breaking models.\foot{This was proven in \RayWK\ for renormalizable K\"aher potentials.  Additional non-renormalizable K\"aher  potential terms, which are present (with unknown coefficients) in the IR-free low-energy duals, contribute to lifting the pseudomoduli.
As we will discuss further in what follows,  such contributions are negligible in ``calculable" models of DSB.} One is the superpartner of the Goldstino, and typically there are many others, corresponding to (at least some of) the moduli of the IR-free theory before turning on the supersymmetry-breaking perturbation.  To definitively determine whether or not supersymmetry is broken requires determining what happens to all of the pseudomoduli in the quantum theory. In the context of DSB in IR-free duals, as we will discuss, pseudomoduli are either  ``good," ``bad," or ``incalculable," depending on their quantum effective potential and how it is generated.

We will distinguish {\it calculable} DSB models, where the demonstration of DSB is under full control, from models where incalculable quantum effects could be important.  In the original models of DSB \AffleckXZ, calculability required that the  fields of the electric theory be far from the origin, i.e.\ $|Q_{elec}|\gg |\Lambda|$, where $\Lambda$ is the strong-coupling scale.   On the other hand, calculable DSB in a low-energy IR-free dual requires  the dual fields to be close to the origin, $|q_{mag}|\ll |\Lambda|$, in order for  unknown higher-dimension operators, which are suppressed by powers of $|\Lambda|$, to be unimportant.     

The condition $|q_{mag}|\ll |\Lambda|$ can be non-trivial to check for the pseudomoduli fields $\Phi$ as it entails computing their quantum effective potential.    A model has calculable DSB only if $V_{eff}(\Phi)$ stabilizes all pseudomoduli 
below the cutoff scale, $|\Phi|\ll |\Lambda|$.    All bets are off if any pseudomodulus has a potential with a runaway\foot{For calculable electric DSB \AffleckXZ,  one instead checks that $V_{tree}$ prevents $Q_{elec}\to \infty$ runaways.} to the cutoff of the low energy theory, $\ev{\Phi}\sim \Lambda$.  In the context of metastable DSB, one must also ensure that no pseudomodulus gives a sliding direction  down to the SUSY vacuum.   Because the low-energy theory is IR free, the lowest non-trivial loop order of  $V_{eff}(\Phi)$ suffices. In the SQCD example,  all pseudomoduli are safely stabilized at one loop in the low-energy theory \ISS.     But in many other potentially interesting generalizations, 
e.g. \refs{\FrancoES\OoguriPJ\ForsteZC\AmaritiVK{--}\GKKS}, some pseudomoduli are unlifted at one loop, so a higher-loop analysis is then required to determine if they have dangerous runaways to $\ev{\Phi}\sim \Lambda$.

We will here consider general aspects of pseudomoduli, and their dynamical lifting 
by $V_{eff}(\Phi)$. This will serve to determine whether pseudomoduli are ``good," "bad," or ``incalculable." Briefly put, we refer to pseudomoduli as ``good" if their quantum effective potential is calculable and robust, stabilizing them within the regime of validity of the IR-free low-energy theory.  The ``bad" pseudomoduli, on the other hand, have a calculable, robust potential, but with a runaway to the cutoff of the low-energy theory.  Finally, the ``incalculable" pseudomoduli are inconclusive, because their
quantum effective potential is not robust against incalculable effects from modes outside of the low-energy theory.

We should note that the ``bad" and ``incalculable" cases can be salvaged by a simple fix, which has already been implemented in models in the literature: one can modify the ultraviolet theory under consideration to give any dangerous pseudomoduli $\Phi_d$ a supersymmetric mass ``by hand." 
This can be done by introducing additional gauge singlets $\Sigma$, coupled to pseudomoduli $\Phi_d$ via $W_{tree}\supset m \Sigma \Phi_d$, as was implemented for the examples in \refs{\FrancoES, \OoguriPJ}.
(This generally introduces additional pseudomoduli, which  need to be examined.)  Alternatively, one can add the term  $W_{tree}\supset m \Phi _d ^2$, as in \refs{\AmaritiVK,\GKKS}.  (This can  introduce new supersymmetric vacua, so the lifetime of any DSB vacua must be re-checked.)   From the perspective of the original UV theory, these are modifications of $W_{tree}$ by some particular higher-dimension operators.  In fact, the recent work \GKKS\ illustrates, in the context of a particular example, how such a modification can turn ``bad" pseudomoduli into a model-building virtue, providing a meta-stable vacuum where R-symmetry is broken spontaneously.

\subsec{Methods and connection to gauge mediation}

In our general analysis, we will find it useful to adapt the language of gauge mediation in order to characterize the coupling of the pseudomoduli to the SUSY-breaking sector.   Recall the idea of gauge mediation (see e.g.\ \GiudiceBP\ for a review and references):  to communicate ``hidden" sector SUSY-breaking to the ``visible" sector (the MSSM or some extension) via loops of ``messenger" fields which couple directly to the SUSY-breaking fields and which are charged under the SM gauge groups. In a wide class of gauge mediated models, the details of the SUSY-breaking sector are irrelevant and the dynamics can be described in terms of a spurion field $X$ that breaks supersymmetry spontaneously through its F-component expectation value, $\ev{X}=M+\theta^2 f$.\foot{We consider $F$-term breaking, where $\varphi$ is a chiral multiplet. As a concrete example, the SUSY-breaking sector can be a generalized   O'Raifeartaigh model, e.g. one like
\eqn\woneis{W_{low}\supset X \varphi ^2+ \phi \varphi ^2+fX,}
where $\phi$ and $\varphi$ contain multiple fields, in representations of a group. This wide class  of 
supersymmetry breaking models, with only cubic and linear terms  (called O'R${}_2$ in \ISSii), includes the original inverse hierarchy model of \WittenKV, and also the rank-condition supersymmetry breaking of \ISS.    Accounting for  $\ev{\varphi}\neq 0$ in these models,  they are of type 1 in the classification of \CheungES.  If $\phi$ and $\varphi$ are charged under a sufficiently strong (but still perturbative) gauge group, some pseudomoduli can have a minimum away from the origin,  spontaneously breaking the (accidental) $U(1)_R$ symmetry of this O'Raifeartaigh sector; see e.g. \refs{\WittenKV,  \DineXT, \ISSii}.
These examples are given only for illustrative purposes.  } The spurion then couples to messengers $\varphi$, $\widetilde\varphi$\ via
\eqn\messparadigm{
W\supset h_X X\varphi \widetilde \varphi,
} 
which gives the messenger scalars SUSY-split masses at tree-level (and the MSSM sfermions soft masses at two loops \gmed).

As we will see, the key point is that in order for the pseudomoduli potential to be calculable, the theory necessarily has some  ``messenger" fields $\varphi$ with SUSY-split tree-level masses. Then the pseudomoduli are analogous to the sfermions of the visible sector, and they will feel the effects of SUSY-breaking through weakly-coupled ``messengers."  Using this language of messengers, we will apply and extend various results and techniques developed for gauge mediation to this new context. 

Each calculable pseudomodulus $\Phi$ is lifted at the loop order given by the number of relevant interactions needed to couple it to the messengers $\varphi$.  (As we will discuss, the pseudomodulus is ``incalculable"
if any of these interactions are power-law irrelevant.)  For some classes of models we will consider, the pseudomoduli are first lifted at three or more loops. Since it would be difficult (to say the least) to directly compute the needed multi-loop effective potential for such cases, we will instead develop here a simpler method, which is related to those of  \refs{\KaplunovskyYX, \GiudiceNI, \AGLR}, and which allows us to determine multi-loop effective potentials in terms of one-loop RG running data. Our method will apply in the regime
\eqn\yrange{m_0\ll |\Phi|\ll |\Lambda|,}
where the pseudomodulus is relatively far from the origin, but still below the cutoff of the low-energy theory.   Using our method, one can easily determine whether calculable pseudomoduli are ``good" or ``bad" in this regime.  Examples suggest that the behavior of the potential in the range \yrange\ is indeed a reliable indicator of whether the pseudomodulus is good or bad.

The analogy between our analysis of pseudomoduli and the standard analysis of gauge mediation is only an analogy, and it is important to stress some differences between the two scenarios:

\lfm{1.} The most obvious difference is that in actual gauge mediation the sfermions only acquire their soft masses through loops involving SM gauge fields, but here the messenger couplings to pseudomoduli are not restricted by flavor considerations. So as we will see in various examples, the pseudomoduli potential can involve both Yukawa and gauge interactions, and it can start at any loop order.

\lfm{2.} The messengers $\varphi$ can have large SUSY-breaking mass splittings (as in \ISS\ where some messengers have $x\equiv |F_X/M^2|=1$).  Then the beautiful methods \refs{\KaplunovskyYX, \GiudiceNI, \AGLR}\ which have been developed for $x\ll 1$,  
to extract multi-loop effects from one-loop data, are not directly applicable.  Nevertheless, we will here discuss a different limit, where similar methods can be employed. 

\lfm{3.} Something that generally does not happen in gauge mediation, but which can easily happen here, is that some messengers can have $\ev{\varphi}\neq 0$ (again, as is the case in \ISS), and this can partly or fully Higgs a hidden-sector gauge group.  (See \EGMS\ for additional discussion of mediation by Higgsed gauge groups.) 

\lfm{4.} Because the low energy theory is assumed to be IR free, all interactions between pseudomoduli $\Phi$ and messengers $\varphi$ are  either marginally irrelevant -- Yukawa couplings or IR free gauge couplings -- or power-law irrelevant.  

\lfm{5.} Being SUSY-breaking mediation in an effective theory, there are unknown 
higher dimension operators, suppressed by powers of $1/|\Lambda|$, which could be potentially dangerous; in the ``good" cases, such operators are unimportant. 

\medskip

\subsec{Summary of the survey of pseudomoduli}

Let us now survey various types of pseudomoduli, and their dynamical lifting. While some of the pseudomoduli may seem rather contrived, all of the ones in this list occur
``naturally" in the effective magnetic description of some strongly-coupled gauge theory. We will study these theories in more detail in section 5, and with examples in section 6.

\bigskip

$\bullet$ {\bf Gauge singlet pseudomoduli, with cubic direct couplings to messengers.}
\eqn\wgenform{W_{low}\supset \Phi_1 \varphi^2+\tilde\Phi_1\varphi\chi,}
where $\Phi_1$ and $\tilde\Phi_1$ are gauge-singlet pseudomoduli, $\varphi$ have SUSY-split tree-level masses, whereas $\chi$ do not.  For light messengers $\varphi$, the $\Phi_1$, $\tilde\Phi _{1}$ fields enter into the one-loop Coleman-Weinberg \ColemanJX\ potential $V_{eff}^{(1)}$, which safely stabilizes such pseudomoduli near the origin.   Such pseudomoduli are ``good," as they have a calculable potential which prevents their vev from sliding to the cutoff.   

$\bullet$ {\bf Gauge singlet pseudomoduli, with cubic indirect couplings to messengers.}
Add to \wgenform\ the term
\eqn\wgenformii{W_{low}\supset \Phi_{2}\chi^2.}
Pseudomoduli like $\Phi _{2}$  are first lifted at two loops, since they couple to the
messengers via $\Phi _{2}\leftrightarrow \chi \leftrightarrow \varphi$, where each $\leftrightarrow$ costs a loop via a Yukawa interaction.   An example realizing such pseudomoduli is SQCD with both massive and massless flavors \FrancoES; the full two-loop effective potential for the analogue of $\Phi _{2}$ in this example was recently explicitly computed in \GiveonWP, and was shown to have monotonically decreasing runaway behavior.   We will here use a simpler analytic method to determine the potential,
using only one-loop data, in the regime of relatively large $\ev{\Phi _{2}}$.  The potential in this regime reveals that such pseudomoduli are ``bad," as they have a calculable runaway potential pushing their vev to the cutoff of the low-energy theory.

$\bullet$ {\bf Higgsing pseudomoduli, gauge-coupled to messengers.} Charged matter fields can lead to pseudomoduli  $\Phi_q$, corresponding to their D-flat expectation values\foot{If there are non-zero D-term expectation values, such fields can be  lifted by tree level or one-loop supersymmetry breaking effects, and thus not be pseudomoduli after all.  We thank N. Seiberg for this comment. }.  For lack of a better name, we call these ``Higgsing" pseudomoduli.   If the messengers $\varphi$ are charged under the same gauge group as the matter $\Phi _q$, the coupling $\Phi_q\leftrightarrow$ gauge $\leftrightarrow\varphi$, leads to a two-loop effective potential (again, each $\leftrightarrow$ coupling costs a loop). In this sense, this type of pseudomodulus is most directly analogous to that of sfermions in ordinary gauge mediation.  The generalization to determine the sfermion soft mass-squared in the 
here-relevant case of Higgsed gauge groups was recently considered in \EGMS.   As we discuss, these pseudomoduli are safely stabilized: their two-loop effective potential pushes them to the origin.  All pseudomoduli from gauge non-singlets have this good, two-loop, stabilizing effect.

$\bullet$ {\bf Saxion-type pseudomoduli.}  These are gauge-singlet fields $\Phi _3$ that couple to the messengers $\varphi$ only via superpotential interactions 
\eqn\wsaxion{W_{low}\supset \Phi _3 ~p^2,}
to charged matter fields $p$, which in turn couple to the $\varphi$ via gauge interactions.  Since 
$\Phi _3$ pseudomoduli couple to the messengers $\varphi$ via $\Phi _3\leftrightarrow p\leftrightarrow$ gauge $ \leftrightarrow \varphi$, they are lifted first by a {\it three-loop} effective potential.  $\Phi _3$ is referred to as ``saxion-type" because of how it enters into the low-energy theory when it gets a large expectation value.  We will here argue that such saxion pseudomoduli are bad, with a destabilized runaway to the cutoff.  This is qualitatively similar (though differing in the details) to the behavior found in \AGLR\ for the saxions in the context of the usual, heavy-messenger scenario of gauge 
mediation of supersymmetry breaking.

$\bullet$ {\bf Gauge messengers.} In the previous examples, we have assumed implicitly that the SUSY-breaking spurion $X$ in \messparadigm\ is a gauge singlet. If instead it is charged under some gauge group, then  the massive gauge fields themselves become messengers. A classic example is the theory of \WittenKV, where gauge messengers arise from the F-term of an adjoint of $SU(5)_{GUT}$. See also \GiudiceNI\ for a discussion of some general aspects of gauge messengers. In such cases, the Higgsing and saxion pseudomoduli couple more directly to the messengers and their potentials are generated at one lower loop order. As we will discuss, the sign of the pseudo-moduli effective potential is reversed in both these cases, meaning that the Higgsing pseudomoduli are destabilized while the saxion pseudomoduli are stabilized by the gauge messengers.

$\bullet$ {\bf Irrelevantly coupled pseudomoduli:}  are gauge-singlet pseudomoduli which 
couple to the SUSY-breaking sector only via power-law irrelevant interactions.  For example, pseudomoduli $\Phi _4$ with direct coupling to messengers and some other fields, 
\eqn\wfour{W_{low}\supset {1\over \Lambda ^{n+m-2}}\Phi _4\varphi  ^n p^m,}
for $n+m\geq 3$.  (As we discuss,  some of these
 interactions can become relevant,  if $m<3$ and $\ev{\varphi}\neq 0$.)  Such pseudomoduli are {\it not} reliably lifted by quantum effects in the low energy theory: the calculable effective potential in the low energy theory is not parametrically larger than incalculable effects of the 
 unknown irrelevant terms in the effective Kahler potential.  All such models are thus inconclusive: whether or not their pseudomoduli are dynamically stabilized depends on the sign of terms which cannot, in principle, be calculated with currently known methods.  

\subsec{Outline}

The outline of this paper is as follows.  The next section discusses general aspects of DSB in IR-free duals.  In section 3, we note that models with power-law irrelevantly coupled pseudomoduli are always inconclusive.  In section 4, we note that there is a limit where even multi-loop pseudomoduli effective potentials can be easily computed from one-loop quantities: this is the limit where the pseudomodulus is far from the origin, as compared with the tree-level mass scale, but still below the cutoff.  In section 5, we use these results to survey which of the above pseudomoduli types are safe, and which have a dangerous runaway to the cutoff.  In section 6, we apply these results to comment on a number of examples.

\newsec{Generalities of DSB in IR free duals}

The low-energy theory is assumed to be an IR-free effective theory with a cutoff scale given by $\Lambda$. The low-energy theory will in general have a variety of mass scales, including the SUSY-breaking scale set by the parameter $f$ in $W_{low}\supset fX$. This mass scale must be well below the cutoff of the low-energy theory, so we define a small parameter  $\epsilon$, given by
\eqn\fepsilon{ \epsilon\equiv {f\over \Lambda ^2}\sim \left({m_0\over \Lambda}\right)^2 \qquad\hbox{with}\quad |\epsilon |\ll 1.}
Calculable IR-free DSB requires such a small parameter $\epsilon$.\foot{The parameter $\epsilon$ is related to a superpotential coupling $\lambda$ of a dual, UV description by $\epsilon = \lambda \Lambda ^{\Delta _{UV}-3}$, where $\Delta _{UV}>1$ is the UV dimension of the composite operator $X$ (and $\Delta _{IR}=1$).  Thus $\epsilon \ll 1$ is natural if  $\Delta _{UV}\geq 3$, so the UV coupling is irrelevant or marginally irrelevant.  If $\Delta _{UV}<3$, the small parameter $\epsilon$ could still be naturalized by additional dynamics \DineGM.}   The mass $m_0$ in \fepsilon\ sets the scale of the tree-level masses in the low-energy theory, as well as the supersymmetry-breaking scale.  
The IR-free low-energy theory has unknown corrections from higher dimension operators, in particular, K\"ahler potential corrections, suppressed by powers of $1/|\Lambda|$.  Such incalculable terms contribute to the  pseudomoduli potentials -- for example
$K_{incalc}\supset c X\bar X\Phi \bar \Phi/|\Lambda|^2$ leads to $V_{incalc}\supset -c |m_0|^4|\Phi |^2/|\Lambda|^2$ with unknown $\CO(1)$ coefficient $c$.   In general, there are incalculable contributions to pseudomoduli potentials  of the form
\eqn\mincalc{ V_{incalc} \sim |m_0|^4 f_{incalc}\left({|\Phi|^2\over \Lambda ^2}\right) \sim |\epsilon|^2\Lambda^4 f_{incalc}\left( |\epsilon|{|\Phi|^2\over m_0^2}\right)
}
where the real analytic function $f_{incalc}$ has a regular Taylor expansion around the origin. On the other hand, the calculable effective potential in the low-energy theory can depend only on $m_0$ and not on $\Lambda$, so 
\eqn\calcVsim{
V_{calc} \sim |m_0|^4 f_{calc}\left({|\Phi|^2\over m_0^2}\right) \sim |\epsilon|^2\Lambda^4 f_{calc}\left({|\Phi|^2\over m_0^2}\right)
}
for some real function $f_{calc}$. The calculable potential is robust against the unknown effects provided that $|\epsilon|\ll 1$ and $|\Phi|\ll |\Lambda|$.

The IR-free low-energy theory can have marginally irrelevant coupling constants, like the Yukawa coupling $h_X$ in $W_{low}\supset h_X X\varphi ^2$, or the gauge coupling $g$ of an IR-free gauge group.  
Such couplings take some fixed, but unknown $\CO (1)$ values at the UV cutoff $|\Lambda|$ of the low-energy theory, $g(|\Lambda|)\sim h(|\Lambda|)\sim \CO (1)$, and then run down to smaller values in the IR.  The running is over a large energy range,  from  
$|\Lambda|$  down to the much lower scale $m_0$ of the tree-level masses, below which the running essentially stops.   The couplings thus run down to small IR values. However, it is important that they are not {\it too} small:
\eqn\ghepsilon{ {g^2\over 16\pi ^2}\sim {h^2 \over 16\pi ^2}\gtrsim (-\ln |\epsilon |)^{-1}\qquad\hbox{so}\qquad |\epsilon| \ll {|g|^2\over 16\pi ^2},\ {|h| ^2\over 16\pi ^2}\ll 1.
}
This ensures  that perturbation theory in the low-energy theory is reliable, with higher-order perturbative effects suppressed as compared with leading-order effects.   A calculable $\ell$-loop mass-term contribution coming from gauge or Yukawa interactions generally has $m^2_{calc, \ell }\sim |\epsilon \Lambda ^2 h^{2\ell +1}|$,  so  pseudomoduli are parametrically lighter, by appropriate powers of the loop-factor \ghepsilon, than the fields which get tree-level masses.  Nevertheless,  for any $\ell$, it follows from \ghepsilon\ that their calculable mass and potential can be robust, 
$m^2_{calc, \ell }\gg m^2_{incalc}$.

Finally, let us remark that non-perturbative effects are insignificant  as long as pseudomoduli are not too far from the origin.  Non-perturbative effects can only become significant if the low-energy theory is driven interacting by a sufficiently large  pseudomodulus expectation value, $|\Phi |>|\Phi _{n.p.}|$. 
Since the low-energy theory is IR free, the scale $\Phi _{n.p.}$ where non-pertubative effects could become important is generally  above the mass scale of the light fields, $|\Phi _{n.p.}|\gg m_0$.   For example, non-perturbative effects are irrelevant for the metastable DSB vacua of \ISS, but the $W_{n.p.}$ eventually becomes important, and leads to the SUSY vacua,  at the scale $\Phi _{n.p.}\sim \Lambda _{low}
\sim \epsilon ^{(N_f-N_c)/N_c}\Lambda$.   In the regime where pseudomoduli are not too far from the origin, $|\Phi|\ll |\Phi _{n.p.}|$, perturbative effects are the most important, and non-pertubative effects
are negligible.  For example, even if non-perturbative effects happen to generate a runaway for a pseudomodulus, the pseudomodulus could still be safely stabilized in the regime $|\Phi|\ll |\Phi _{n.p.}|$ by the larger perturbative effects  there (see \EssigXK\ for an example of this).

\newsec{Irrelevantly coupled pseudomoduli are always inconclusive}

In this section, we discuss power-law irrelevantly coupled pseudomoduli, like $\Phi _4$  in \wfour.  Consider the least irrelevant example, 
\eqn\wexamp{W_{low}\supset {1\over \Lambda }\Phi _4\widetilde \Phi _4 \varphi ^2.}
This term can potentially become relevant, if the DSB vacuum has $\ev{\varphi}  \neq 0$.  For example, we could have $\ev{\varphi ^2}\sim \epsilon \Lambda ^2/h$ as in \ISS,  where we have included an IR-free coupling constant $h$  \ghepsilon.  (We will illustrate this with a concrete example in section 6.5.)  In this case, (some components of) $\Phi _4$ are not pseudomoduli after all -- they get a tree-level  supersymmetric mass $m_{calc}\sim \ev{\varphi ^2}/\Lambda \sim \epsilon \Lambda/h$.  Comparing this $m_{calc}$ with the unknown mass contribution \mincalc, $m^2_{incalc}\sim |\epsilon \Lambda |^2$, we see that the calculable tree-level masses are here just barely larger, and thus just barely robust, thanks to the $h^{-2}\sim -\ln |\epsilon |$ enhancement of $m_{calc}^2$.  

When some fields like $\Phi _4$ in \wexamp\ are pseudomoduli, however, their $m_{calc}^2$ comes with additional loop factors of the IR-free couplings, so $m^2_{calc}\sim  |h^{\ell -1}\epsilon \Lambda |^2$, which for any loop order $\ell\geq 1$ is not robust against the  incalculable contributions \mincalc. 
Pseudomoduli $\Phi _4$ with couplings which are more irrelevant than \wexamp\ have even smaller $m_{calc}^2$.   The conclusion is that the effective potentials for power-law irrelevantly coupled pseudomoduli can never be reliably computed in the low-energy effective field theory.  It is impossible to determine  whether or not such pseudomoduli are safely stabilized at a vacuum within the regime of validity of the low-energy theory, with expectation values properly below its cutoff $\Lambda$.   Even if the low-energy theory appears to break supersymmetry at tree-level, 
supersymmetry might not be broken after all, if there is no static DSB vacuum within the regime of validity of the low-energy theory.   We refer to such pseudomoduli and theories as {\it incalculable}. 

This point afflicts and renders as inconclusive many potential examples of (perhaps metastable) DSB via IR-free duality dynamics; for example, all of the duality examples of \refs{\KutasovVE\KutasovNP-\KutasovSS}, and the many other similar generalizations.   All such examples have incalculable pseudomoduli.  Thus {\it none} of these examples can have reliably calculable metastable DSB -- they are all inconclusive. As discussed in the introduction, one can still modify the UV theory by hand to give  tree-level masses to the incalculable pseudomoduli, as in 
\refs{\OoguriPJ, \AmaritiVK}, for example.

\newsec{A regime where the pseudomodulus'  potential follows simply from running}

The effective potential for pseudomoduli which are lifted at one loop is easily computed from the
expression for $V_{eff}^{(1)}$ of \ColemanJX.  For  pseudomoduli which are first lifted at two loops, one can, in principle, use the expression for $V_{eff}^{(2)}$ in \MartinVX, though in practice this can be quite technically involved   -- see \GiveonWP\ for an example and some methods.  And pseudomoduli like the saxion, which is lifted first at three loops, would require extensive work in order to evaluate $V_{eff}^{(3)}$.   

Here we note that there is a useful regime where all of the general higher-loop effective potentials can be easily determined by one-loop quantities, through a generalization of the wavefunction renormalization methods of  \refs{\GiudiceNI,\AGLR}. The regime of interest is where the pseudomodulus $\Phi$ is relatively far from the origin but still within the validity of the low-energy effective theory, i.e.\foot{We ignore any factors of coupling constants which could be multiplying $\Phi$ in this section, since they will be irrelevant to the discussion.}
\eqn\yrange{m_0\ll |\Phi|\ll |\Lambda|,}
where $m_0$ is the typical mass scale of the light fields in the low-energy theory, or equivalently the scale at which SUSY is broken. (We also assume that 
 $|\Phi|\ll |\Phi_{n.p.}|$, so non-perturbative effects are negligible.) It can be useful to know the potential in this regime, since if it increases with $|\Phi|$, then we can be sure that the pseudomodulus must be stabilized {\it somewhere} along its flat direction. On the other hand, if the potential decreases in the
 range \yrange,  then this is evidence for runaway behavior -- although, from this computation alone, one cannot rule out the possibility that there is a local minimum of the effective potential near the origin.

To compute the effective potential in the regime \yrange, we use the fact that the pseudomoduli only couple to the other fields in the theory linearly. Thus at large $\Phi$, all that happens is some fields of the low-energy theory get masses $\sim \Phi$. Moreover, since $|\Phi|\gg m_0$, these massive fields are approximately supersymmetric. So to a good approximation, integrating them out yields an approximately supersymmetric effective theory below the $\Phi$ scale, where the only dependence on $\Phi$ comes from threshold effects in the effective K\"ahler potential. If we assume for simplicity that a single field $X$ has nonzero F-term vev, $F_X=f\neq 0$, then the Wilsonian effective action below the scale $\Phi$ takes the form
\eqn\efftheorygen{
K_{eff} = Z_X(Q;|\Phi|)X^\dagger X + \dots,\qquad W_{eff} = f X + \dots
}
where $Q$ is the RG scale and $Z_{X}$ is $X$'s wavefunction renormalization. In the regime \yrange, the leading-log-enhanced dependence of $Z_X$ on $\Phi$ is determined using only one-loop supersymmetric RGEs. Then using this in computing the {\it tree-level} vacuum energy in the effective theory gives the leading approximation to the effective potential for $\Phi$:
\eqn\Vyapproxi{
V_{eff}(\Phi)\approx  |f|^2 Z_X(m_0;|\Phi|)^{-1}
}
in the regime \yrange. 

We will find it convenient to introduce the notation  $\Omega_X=-{1\over2}\log Z_X$ so that the anomalous dimension of $X$ is given by
\eqn\anomdim{
\gamma_X ={d\Omega_X\over dt}
}
where $t=\log {Q\over m_0}$ is the RG time. Then \Vyapproxi\ becomes
\eqn\Vyapproxibec{
V_{eff}(\Phi)\approx |f|^2e^{2\Omega_X(m_0;|\Phi|)}
}
The details of the calculation of $\Omega_X$ are contained in the appendix.  The upshot is that the lowest-order leading-log contribution to $\Omega_X$ is given by
\eqn\ZXupshot{
\Omega_X(m_0;|\Phi|) = const. - {1\over n!}\Delta \Omega_X^{(n)}(-t_\Phi)^n+ \CO(\kappa^{n+1})
}
where $t_\Phi \equiv \log {|\Phi |\over m_0}$ and $const.$ refers to the $|\Phi|$ independent part of the wavefunction; $\kappa$ is the loop-counting parameter (like $\kappa _h= h^2/16\pi ^2$ or $\kappa _g=g^2/16\pi ^2$); and $\Delta \Omega_X^{(n)}$ is short for 
\eqn\ZXdisc{
\Delta \Omega_X^{(n)}\equiv {d^n\Omega_X\over dt^n}\Big|^{t_\Phi^+}_{t_\Phi^-}
}
i.e.\ the discontinuity at $t=t_\Phi$ in the $n$th derivative of $\Omega_X$ with respect to RG time.  Each derivative of $\Omega_X$ with respect to RG time brings down a factor of the loop-counting parameter, so $\Delta \Omega_X^{(n)}\sim\CO(\kappa^n)$ if one uses the one-loop anomalous dimension in \anomdim.  Higher-loop corrections to the anomalous dimension add additional factors of $\kappa$ and do not contribute at lowest leading-log order, so in the following all anomalous dimensions and beta functions will implicitly be one-loop quantities in order to simplify the notation.  

Explicitly, we have for the first few values of $n$:
\eqn\ZXdiscare{\Delta \Omega _X^{(1)}=\Delta \gamma _X, \qquad \Delta \Omega _X^{(2)}=\sum _I {\partial \gamma _X\over \partial g^I}\Delta \beta ^I, \qquad \Delta \Omega _X^{(3)}=\sum _{I,J}{\partial \gamma _X\over \partial g^I}{\partial \beta ^I\over \partial g ^J}\Delta \beta ^J,}
where $\beta ^I=dg^I/dt$ and $\Delta$ refers to the discontinuity across the $\Phi$ threshold as in \ZXdisc. Note that these formulas for $\Delta \Omega ^{(n)}_X$ assume that the lower order $\Delta \Omega _X^{(m<n)}$ vanish. This is why, for example,  we have not written a contribution to $\Delta\Omega_X^{(3)}$ from ${\partial^2 \gamma_X\over\partial g^I\partial g^J} \beta^I\Delta\beta^J$, since if such a term were non-zero it would've already contributed to $\Delta\Omega_X^{(2)}$  as well.

In any event, substituting \ZXupshot\ into \Vyapproxibec, we obtain at the lowest leading-log order:
\eqn\Vyapproxii{
 V_{eff}(\Phi) \approx const. -{2\over n!}V_0 \Delta\Omega_X^{(n)}\left(-\log {|\Phi|\over m_0}\right)^n
 }
where $V_0=|f|^2$ is the tree-level vacuum energy. In this way, the $n$-loop leading-log potential is completely determined by one-loop quantities.\foot{As will be clear in the examples, the order $n$ of the leading-log effective potential approximation \Vyapproxii\ indeed agrees with the expected loop order, given by the number of interactions needed to couple the pseudomodulus $\Phi$ to some messengers with SUSY-split masses. Thus we can be confident that the approximations used to obtain \Vyapproxii\ are indeed capturing the dominant term in the effective potential (in the regime \yrange).} The sign of the coefficient of the leading-log term then determines whether the pseudomodulus $\Phi$ is ``good" or ``bad" -- in short, $\Phi$ is
\eqn\summary{\cases{\hbox{``good"}& if $(-1)^{n+1}\Delta \Omega _X^{(n)}>0$\cr 
\hbox{``bad"} & if $(-1)^{n+1}\Delta \Omega _X^{(n)}<0$,}}
where the loop order $n$ is the lowest number for which $\Delta \Omega _X^{(n)}\neq 0$.

It is trivial to generalize to the case where multiple fields $X_i$ have non-zero F-terms. Then the leading effective potential for a pseudomodulius $\Phi$ in 
the range \yrange\ is given by:
\eqn\Vyapproxis{V_{eff}\approx \sum _i |F_{X_i}|^2 (Z_{X_i}(m_0; \Phi))^{-1}\equiv \sum _i |F_{X_i}|^2e^{2\Omega _{X_i}(m_0;|\Phi|)}~,}
and each term in the sum can be approximated as in \Vyapproxii,
\eqn\Yyapproxiix{|F_{X_i}|^2( Z_{X_i}(m_0;\Phi))^{-1}\approx const. -{2\over n_i!}|F_{X_i}|^2 \Delta \Omega _{X_i}^{(n_i)}\left(-\log {|\Phi|\over m_0}\right)^{n_i},}
where $\Delta \Omega _{X_i}^{(n_i)}$ is defined as in \ZXdisc.
Then the potential \Vyapproxis\ is approximated by keeping only those terms $i$ with the lowest loop order, i.e.\ the smallest value of $n_i$.    In the next section, we will apply \Vyapproxii\ and \Yyapproxiix\ to the cases of interest.

Finally, let us make a few comments on the various corrections to \Vyapproxi.

\lfm{1.} Finite effects cannot be captured by the one-loop RGEs, but they are clearly subleading compared to the large logarithms. 

\lfm{2.} Loop effects in the effective theory below the scale $\Phi$ only depend on $\Phi$ through the wavefunctions, so they are clearly subleading as well.

\lfm{3.} In order for \Vyapproxii\ to be the dominant term in the effective potential at large $\Phi$, all of the one-loop anomalous dimensions and beta functions must be nonzero. In all the examples we study, this will indeed be the case. If this condition is not satisfied, and some anomalous dimensions or beta functions vanish, then subleading logarithms or even finite effects at a lower loop order could be larger than the effect shown in \Vyapproxii. 

\lfm{4.} The mistake we are making in assuming that the theory is a supersymmetric effective theory described by \efftheorygen\ comes in the form of terms that are higher order in $f/|\Phi|^2$. In the regime \yrange\ these are clearly negligible compared to the log-enhanced effects.

\medskip

\newsec{Surveying the pseudomoduli}

We here discuss each of the pseudomoduli types mentioned in the introduction. In each case, we will apply our general result \Vyapproxii\ for the leading-log effective potential in the regime \yrange.

\subsec{Gauge singlet pseudomoduli, with Yukawa coupling to messengers}

Let us rewrite \wgenform\ and \wgenformii\ with Yukawa couplings reintroduced,
\eqn\hwgenform{W_{low}\supset h(\Phi_1\varphi^2  + \tilde\Phi_1\varphi\chi + \Phi_2\chi^2)}
 Here the  $\varphi$ have SUSY-split tree-level masses, whereas $\chi$ do not. We take all the Yukawa couplings equal just for simplicity, so the loop-counting parameter is uniformly given by
\eqn\kappadef{
\Kappa = {h^2\over 16\pi^2}.
}
The $\varphi$ couple to fields $X$ (which can differ from the $\Phi _1$ fields) with $F_X\neq 0$ and, as in \messparadigm, we distinguish that Yukawa coupling as $h_X$. When $\Phi=\Phi _{1}$ or $\Phi=\tilde\Phi_1$ is large, the anomalous dimension of $X$ is discontinuous at the scale $|h\Phi|$ 
since the number of messengers is reduced below this scale:
  \eqn\phionead{\gamma _X=n_\varphi \kappa _{h_X}, \qquad \hbox{so}\qquad \Delta\gamma _{X}=\Delta n_{\varphi}\kappa_{h_X},
  }
where $\Delta n_{\varphi} = n_{\varphi}-n_{\varphi}'>0$ is the discontinuity in the number of messenger fields above and below $|h\Phi|$. Thus the effective potential can be determined from the $n=1$ case of \Vyapproxii, leading to the well-known result
\eqn\vxoknownlim{V_{eff}\approx 2V_0\Delta \gamma _X \ln {|h\Phi|\over m_0}.}
Since $V_0>0$ and $\Delta\gamma_X>0$,  such pseudomoduli are safely stabilized below the cutoff of the IR-free dual by the one-loop effective potential. (Depending on the specifics of the model, they could be stabilized either at the origin or away from it. For instance, the second term in \hwgenform\ is analogous to the messenger-matter mixing considered in \DineXK; as shown there, it can lead to a negative mass-squared for $\tilde\Phi_{1}$ at the origin, so the minimum is elsewhere.) 

On the other hand, pseudomoduli like $\Phi _{2}$ in \hwgenform\ require two loops to couple to the messengers, via $\Phi _{2}\leftrightarrow \chi \leftrightarrow \varphi$.   In principle, the two-loop effective potential for $\Phi _{2}$ can be obtained from the general expressions in \MartinVX; this was recently carried out in the context of SQCD with massive and massless flavors in \GiveonWP.  Here we simply consider the potential for $\Phi=\Phi _{2}$ in the range \yrange.   The $\chi$ fields in \hwgenform\ then get a large mass $\sim |\Phi_{2}|$, and can be integrated out at lower scales. This does not cause a discontinuity in $\gamma_X$ but it does affect its first derivative with respect to RG time. In more detail, we have
\eqn\twoloopif{
\Delta \Omega_X^{(2)} = \Delta \left({d \gamma_X\over d t}\right)
={\partial \gamma_X\over \partial h_X}\Delta \beta _{h_X} = 4n_{\varphi}\kappa _{h_X}\Delta\gamma_{\varphi},
}
where in the last equality we have used $\gamma_X=n_{\varphi}\kappa _{h_X}$ and $\beta_{h_X}=h_{X}(\gamma_X+2\gamma_{\varphi})$. Finally, taking
$\Delta \gamma_{\varphi} =n_{\chi} \Kappa$ with $n_{\chi}$ being the number of $\chi$ fields which got a mass from $\Phi_{2}$, this becomes
\eqn\twoloopifii{
 \Delta\Omega_X^{(2)}= 4n_{\varphi} n_{\chi}\kappa _{h_X}\Kappa
}
Substituting into \Vyapproxii, we obtain 
\eqn\vphitt{V_{eff}(\Phi _{2})\approx  -V_0 \Kappa \kappa _{h_X}n_{\varphi} n_{\chi}\left(\ln {|h \Phi _{2}|^2\over m_0^2}\right)^2,}
Therefore, the two-loop potential for $\Phi_{2}$ reveals a destabilized runaway.

\subsec{Higgsing pseudomoduli}

As described in the introduction, these pseudomoduli come from expectation values of
matter fields $\Phi_q$ charged under a gauge group in which the messengers $\varphi$ also transform.  The physics is quite different depending on whether  the field(s) $X$ with $F_X\neq 0$ are neutral or charged under the gauge group.
For charged $F_X$, the gauge fields are  ``gauge messengers" and can lift Higgsing pseudomoduli at one loop. We will discuss this case more in section 5.4 and present an example of it in section 6.2. In this subsection we will focus on the case of neutral $F_X$, where Higgsing pseudomoduli are instead lifted at two loops.

 We consider the limit of large  $\Phi=\Phi_q$, in the range \yrange.  Suppose the messengers $\varphi$ decompose into fields $\varphi_i$ transforming in irreducible representations $r_{\varphi_i}$ of $G$. Above the scale $\Phi_q$, the gauge coupling contribution to the one-loop anomalous dimension of the messengers is given by
\eqn\gcanomdim{
\gamma _{\varphi_i}\supset -2c(r_{\varphi_i})\kappa_g
}
with $\kappa_g=g^2/16\pi^2$ being the loop-counting parameter for gauge coupling $g$ and $c(r_{\varphi_i})$ being the quadratic Casimir invariant. (In general $c(r)=T(r)|G|/|r|$; so $c({\rm fund})={N^2-1\over 2N}$ and $c({\rm adj})=N$ for $SU(N)$.)
 
 The gauge group is (partially or fully) Higgsed down to $G'\subset G$ at the mass scale $|\Phi_q|$, and that affects the anomalous dimensions of the messengers.  Below the Higgsing scale we can decompose the messengers into fields $\varphi_i'$ transforming as $G'$ irreps $r_{\varphi_i'}$. Each field $\varphi_i'$ then has an anomalous dimension given by \gcanomdim\ with $\varphi_i\to\varphi_i'$.
 
As in the previous subsection, \twoloopif\ gives the leading contribution to the effective potential, but instead of \twoloopifii\ we have
\eqn\twoloopifg{
\Delta \Omega_X^{(2)} =  4\kappa _{h_X}\Delta\left(\sum_i |r_{\varphi_i}| \gamma_{\varphi_i}\right)
= -8\kappa _{h_X}\kappa_g \Delta\left( \sum_i |r_{\varphi_i}|c(r_{\varphi_i})\right)
}
Since  $|r_{\varphi_i}|c(r_{\varphi_i}) = T(r_{\varphi_i})|G|$ and the index $T$ is additive, \twoloopifg\ becomes simply
\eqn\twoloopifgii{
\Delta \Omega_X^{(2)} = -8\kappa _{h_X}\kappa_g T(r_\varphi)(|G|-|G|')
} 
Finally, substituting into \Vyapproxii, we have 
\eqn\vtwoex{V_{eff}(\Phi_q)\approx  2V_0\kappa _{h_X}\kappa_gT(r_\varphi)(|G|-|G|')\left(\ln {|\Phi_q|^2\over m_0^2}\right)^2.} 
which implies (since $|G|'<|G|$) that the potential always has a stabilizing effect on the pseudomodulus field. This is analogous to the two-loop potentials for D-flat directions found in \refs{\deGouveaTN, \AGLR}\foot{Indeed the general gauge mediation
\MeadeWD\ effective potential for Higgsing pseudomoduli is 
\IScomments
\eqn\veffcom{V_{eff}(m_W^2)={g^2\over 2}\int {d^4p\over (2\pi )^4}\Tr \left({p^2\over p^2+m_W^2}\right)(3\widetilde C_1-4\widetilde C_{1/2}+\widetilde C_0),}
where $\Tr $ sums over vector bosons, with mass matrix $m_W^2$.  For sfermions $q_f$ in reps $r_{q_f}$ we have  $(m_W^2)^{AB}=\sum _f \Tr (T_{r_{q_f}}^{(A}q_f^\dagger T_{r_{q_f}}^{B)}q_f)$.  We are here interested in the case of weakly coupled messengers $\varphi$, so $C_a(p^2/M^2)$ is $T(r_\varphi)$ times the expressions quoted in \MeadeWD.  Expanding \veffcom\ for $q_f$ near the origin gives the sfermion soft masses $m^2_{q_f}\sim m_0^2g^4 >0$,  including the group theory factor $c(r_{q_f})$.  
As was recently analyzed in \EGMS, 
there can be some numerical differences in the coefficient of $m_{q_f}^2$, as compared with the usual gauge mediation scenario,  because the messengers can have  $\ev{\varphi}\sim m_0\neq 0$, Higgsing the gauge group (as in the SQCD example of \ISS).   Approximating \veffcom\ for some $Q_f\sim \Phi _q$ far from the origin also indeed yields  \vtwoex.}.

\subsec{Saxion-type pseudomoduli}

Consider gauge singlet ``saxion" pseudomoduli $\Phi _3$, coupling to charged matter $p$ via
\eqn\wsaxion{W_{low}\supset h\Phi _3 p^2,}
The $\Phi _3$ potential is first generated at three loops, since they couple to the messengers only via $\Phi _3\leftrightarrow p\leftrightarrow {\rm gauge}\leftrightarrow \varphi$.  In the range \yrange\ of large $\Phi _3$, the saxion effective potential can be determined using \Vyapproxii\ with $n=3$:
\eqn\threeloopif{
\Delta \Omega_X^{(3)} =\Delta \left({d^2\gamma _X\over d t^2}\right)= {\partial \gamma _X\over \partial h_X}{\partial \beta _{h_X}\over \partial g}\Delta \beta _g .
}
This is nonzero, since the $p$ fields are massive and can be integrated out at the scale $h\Phi_3$, which affects the beta function of the gauge coupling below the $h\Phi_3$ threshold:
\eqn\betavar{\Delta \beta _g=-{g^3\over 16\pi ^2}(b-b')=-g\kappa_g(b-b'),}
where e.g. $b=3N-N_f$ for an $SU(N)$ gauge theory with $N_f$ flavors. 
Substituting this into \threeloopif\ and using $\gamma_X=n_\varphi\kappa _{h_X}$, $\beta_{h_X}=h_X(\gamma_X+2\gamma_\varphi)$ and $\gamma_\varphi \supset -2c(r_\varphi)\kappa_g$, this becomes
\eqn\threeloopifii{
 \Delta\Omega_X^{(3)} = 16\kappa _{h_X}\kappa_g^2n_\varphi c(r_\varphi)(b-b')
 }
Then it follows from \Vyapproxii\ that
\eqn\vthreeex{
 V_{eff}(\Phi_3)\approx {2\over 3} V_0 \kappa_{h_X}\kappa_g^2 n_\varphi c(r_\varphi)(b-b')
 \left(\ln {|h\Phi_3|^2\over m_0^2}\right )^3.
 }
Because giving mass to some matter makes the IR group more strongly interacting, we here have  $b-b'=-T_2(r_p)<0$.  With this sign in \vthreeex, we  conclude that saxion-type pseudomoduli $\Phi _3$ always have a destabilizing runaway potential, generated at three loops, at least in the range \yrange.

Let us make two comments on this result. First, note that the change in the beta function \betavar\ is accounted for by an added term in the low-energy theory, 
\eqn\wlowsax{W_{low}\supset  -{1\over 64\pi ^2}(b-b')\int d^2\theta \ln (\Phi_3) W_\alpha ^2,}
which is the only way that $\Phi _3$ enters into the low-energy theory in this limit.  This is the reason for calling such pseudomoduli saxions: the phase of $\Phi _3$ enters the low-energy theory like an axion, and $\ln |\Phi _3|$ is its saxion superpartner.

Second, in the standard gauge-mediation setup, it can be shown  \AGLR\ that  saxion pseudomoduli $\Phi _3$ also have destabilizing, tachyonic $m_3^2<0$ near the origin, in analogy with 
how the top Yukawa and $m_t^2=\CO (\alpha _s^2)$ can lead to electroweak breaking $m_H^2<0$.  
This argument, however, relies on the large separation of scales of a high messenger scale $m_\varphi$: one starts at the high scale $m_\varphi$ with the two-loop supersymmetry breaking $m_p^2|_{\mu =m_{\varphi}}>0$, and $m_3^2|_{\mu =m_\varphi}\approx 0$.  Then $m_3^2|_{\mu \ll m_{\varphi}}<0$ follows from RG running the  one-loop $h_3^2$ Yukawa contribution to the running $m_3^2$ and $m_p^2$.   But in our examples of interest, there are light messengers, no such large range of running, and  the finite terms in the effective potential cannot be neglected in computing the three-loop potential for $\Phi _3$ near the origin. 
So here a full-fledged explicit calculation of the three-loop effective potential for $\Phi _3$ would generally be required to determine if there is any (metastable) local minimum near the origin. In any case, we have argued for the destabilized runaway farther from the origin.

\subsec{Modifications when there are gauge messengers}

Finally, let us discuss models with gauge messengers, i.e.\ models where the gauge multiplets have tree-level, SUSY-split masses.   Gauge messengers occur if any charged matter field has a non-zero, tree-level F-term.
The methods discussed in section 4 can be applied, with only trivial modification,  to determine the effective potential for pseudomoduli when there are gauge messengers.  Though the methods are the same, the physics with gauge messengers is quite different.     Higgsing and saxion-type pseudomoduli are lifted at one fewer loop order by gauge messengers.  Correspondingly, the sign of the leading-log effective potential for these pseudomoduli can be opposite from cases without gauge messengers.  
So Higgsing pseudomoduli can become destabilized at one loop, and saxion  pseudomoduli can become stabilized at two loops.

Let us first consider Higgsing-type pseudomoduli.  With gauge messengers, 
Higgsing pseudomoduli $\Phi _q$ are
lifted at 1-loop.  This can be seen simply from \Vyapproxis\ and \Yyapproxiix\ (we are allowing for the possibility that there are multiple fields $X_i$ with F-terms, transforming in different representations of the gauge group). That is, charged fields $X_i$ have a discontinunity in their 1-loop anomalous dimensions when the gauge group is Higgsed at the scale $|g\Phi _q|$, 
\eqn\gmesshiggs{\Delta \Omega _{X_i}^{(1)}=\Delta \gamma _{X_i}=-2\kappa _g\Delta c(r_{X_i}).}  
Here $\Delta c(r_{X_i})=c(r_{X_i})|_{G'}^{G}$, where $G$ denotes the group above the Higgsing scale, and $G'$ denotes that below, and in general $\Delta c(r_{X_i})>0$.
As in \vxoknownlim, the potential in the large $|\Phi|$ regime is then approximated by
\eqn\vgmhiggs{V_{eff}(\Phi)\approx -2\sum _i |F_{X_i}|^2\kappa _g \Delta c(r_{X_i})\ln {|g\Phi|\over m_0}.}
Because $\Delta c(r_{X_i})>0$, we see that Higgsing pseudomoduli are generally unstable at 1-loop in theories with gauge messengers.  (One exception is if the pseudomodulus $\Phi$ is itself one of the fields $X_i$ with non-zero F-term, which also couples to matter messengers $\varphi$; in this case, there can be also a positive contribution to $\Delta \gamma _{X_i}$ as in \WittenKV.)

Let us now consider saxion-type pseudomoduli, like $\Phi _3$ in \wsaxion.  When such pseudomoduli are in the range \yrange, the change at the scale $\Phi _3$ is 
\eqn\gmesssax{\Delta \Omega _{X_i}^{(2)}=\Delta\left( {d\over dt}(-2c(r_i)\kappa _g)\right)= -{4c(r_i)\kappa_g\over g}\Delta\beta_g = +4\kappa_g^2c(r_i)(b-b').}
where in the last equation we have used \betavar. Then \Yyapproxiix\ gives
\eqn\gmesssaxv{V_{eff}(\Phi _3)\approx \sum _i 4|F_{X_i}|^2 \kappa _g^2 c(r_i)(b'-b)\left(\log {|\Phi|\over m_0}\right)^2.}
 Because $b'-b>0$ in this case (as discussed after \vthreeex), the 2-loop potential stabilizes the saxion.  This has the opposite sign of \vthreeex, and appears at one fewer loop order.  So gauge messengers can stabilize saxions.  

\newsec{Examples}

\subsec{Warmup: SQCD with massive flavors \ISS}

To illustrate our method, and set up the notation for following examples, let us briefly review
the metastable DSB theory based on SQCD with massive flavors \ISS.  The UV theory is $SU(N_c)$ SQCD, with $N_f$ in the IR free-magnetic range, $N_c<N_f<{3\over 2}N_c$.  The $N_f$ flavors have a small mass $m_Q$, which for simplicity we take to be the same for all flavors, 
 so there is a global $SU(N_f)\times U(1)_B\cong U(N_f) $ symmetry.  The low-energy theory is given by the IR free $SU(N=N_f-N_c)$ dual, with 
\eqn\wsqcd{W=h\Tr \Phi \varphi \widetilde \varphi - h\mu ^2\Tr \Phi.}  
 The mass scale is given in terms of UV data as $\mu ^2\equiv(-1)^{N_c /N_f}m\Lambda$ (see \refs{\ISS, \IntriligatorCP}\ for discussion of these factors).  This theory has 
metastable SUSY breaking vacua, given at tree level by  
\eqn\phivarphipm{\Phi = \pmatrix{0&0\cr 0&\widehat \Phi }, \quad \varphi = \pmatrix{\widehat \varphi \cr 0}, \quad \widetilde \varphi=\pmatrix{\widetilde {\widehat \varphi} & 0}, \quad\hbox{with}\quad \widetilde {\widehat \varphi} \widehat \varphi  = \mu ^2\unit _N,}
with $\widehat \Phi $ an $(N_f-N)^2=N_c^2$ matrix of pseudomoduli.  SUSY is broken by 
\eqn\Fvevs{
F_\Phi =\pmatrix{ 0 & 0 \cr 0 & f\unit_{N_f-N}}, \qquad \hbox{with}\qquad f^\dagger  =h\mu ^2\ne 0.
} 
The vacuum energy is $V_0=(N_f-N)|h^2\mu ^4|$, and the tree-level mass scale is $m_0=h\mu$.  

The pseudomoduli are all lifted at one loop, as is evident from the fact that they have direct superpotential coupling to the messengers $\varphi$.  The 1-loop effective potential for all values
of the pseudomoduli was computed in \ISS, and it was noted that the metastable DSB vacua are at 
$\widehat \Phi =0$, $\widehat \varphi = \widetilde {\widehat \varphi}=\mu \unit _N$.  The gauge and global symmetry group is broken in these vacua as  $SU(N)\times U(N_f)\to SU(N)_D\times U(N_f-N)$. 

When the pseudomoduli $\widehat \Phi$ in \phivarphipm\ are in the range \yrange, we can alternatively use the  result \vxoknownlim\ to easily determine the form of the effective potential.  Taking e.g. $\widehat \Phi$ in \phivarphipm\ to have $N_f-N$ independent and large diagonal entries, each of which has  $\gamma _\Phi =N \kappa _h $, and then accounting for the flavor structure, immediately yields 
\eqn\vsqcdex{V_{eff}(\widehat \Phi)\approx |f|^2 N\kappa _h\Tr  \ln {\widehat \Phi ^\dagger \widehat \Phi\over |\mu |^2}.} 
The rising potential \vsqcdex\ gives a quick and indeed reliable indication that the $\widehat \Phi$ pseudomoduli are good.   (As discussed in \ISS, for sufficiently large $\Phi$ the non-pertubative $W_{dyn}$ becomes important and the potential eventually slopes down to the supersymmetric vacua.  This
happens past the far, but still perturbative regime of \vsqcdex.)

\subsec{Example with a two-loop runaway: SQCD with both massive and massless flavors}

We again take the UV theory to be $SU(N_c)$ SQCD with $N_f$ flavors in the free-magnetic phase.   But here
we take only $N_{f1}$ of the flavors to be massive, leaving $N_{f0}$ massless flavors.  For simplicity, we will give the $N_{f1}$ flavors equal mass, $m$.  
In the IR, the superpotential is then of the general form \wgenform.  In particular, for magnetic fields (we use the notation of \GKKS) 
\eqn\Phivarphiare{\Phi = \pmatrix{\Phi _{11} &\Phi _{10}\cr \Phi _{01}&\Phi _{00}} 
, \qquad \varphi=\pmatrix{\varphi_1 \cr \varphi_0}
, \qquad \widetilde\varphi=\pmatrix{\widetilde \varphi_1& \widetilde \varphi_0},}
we have
\eqn\wsqcdmml{W=h\Tr \Phi _{ij}\varphi_j \widetilde \varphi_i -h\mu ^2\Tr \Phi _{11} ,}
where $\Phi _{ij}$ is an $N_{fi}\times N_{fj}$ matrix and $\varphi_i$ and $\widetilde\varphi_i^T$ are $N_{fi}\times N$ matrices.

The first two terms in \wsqcdmml\ lead to rank-condition supersymmetry breaking when $N_{f1}-N=N_c-N_{f0}>0$, with $V_0=(N_c-N_{f0})|h\mu ^2|^2$.  All fields other than $\Phi _{00}$ get masses at tree-level or one-loop level.  This was observed in  \FrancoES, where it was also shown that $\Phi _{00}$ has a non-perturbative runaway coming 
from a dynamically generated superpotential.  This is not yet fatal, as there is a range of the pseudomoduli space where the non-perturbative effects are negligible in comparison with higher-loop perturbative effects. But it was recently shown in \GiveonWP\ that higher-loop perturbative effects also lead to a potential with a runaway to large $\Phi _{00}$.

Let us now use the formalism developed in the previous two sections to demonstrate this two-loop runaway behavior in the range \yrange.   We identify $\Phi_{00}$ with the $\Phi_2$ pseudomodulus of section 5.1.    
 The fields $\Phi _{11}$ have F-terms given by \Fvevs,  replacing $N_f$ with $N_{f1}$, so they play the role of $X$.    The fields $\varphi_1$, $\tilde\varphi_1$ then play the role of the messengers $\varphi$, while the fields $\varphi_0$, $\tilde\varphi_0$ are analogues of $\chi$ in \wgenform.
 
 To simplify the flavor index structure, let us take $\ev{\Phi _{00}}\propto\unit _a\oplus0_{N_{f0}-a}$, for an integer $a$ between 1 and $N_{f0}$.  Below the $\vev{\Phi_{00}}$ threshold, the $a$ flavor components of $\varphi_0$ get a mass and no longer contribute to $\gamma _{\varphi}$, leading to $\Delta \gamma _{\varphi}=a h^2/16\pi^2\equiv a \Kappa$.  As in \twoloopif, we then find $\Delta\Omega_{\Phi_{11}}^{(2)}=4Na\Kappa$.  Accounting for the $SU(N_{f0})_L\times SU(N_{f0})_R$ flavor symmetry,
the effective potential in this regime is found to be 
\eqn\vsqcdmassless{V_{eff}(\Phi _{00})\approx -V_0N\Kappa^2\Tr \left(\ln {|\Phi _{00}|^2 \over |\mu |^2}\right)^2.}
The sign indicates that the $\Phi _{00}$ pseudomoduli have a two-loop runaway to the cutoff of the low-energy theory, and this model thus does not have a calculable metastable DSB vacuum.  

This runaway can be lifted in a variety of ways. As noted in \FrancoES, one way is to add singlets, $\Sigma$, to the UV theory with a marginal superpotential coupling to $\Phi _{00}$, $W_{tree}\supset m\Tr\Phi _{00}\Sigma$.   In this modified theory, all pseudomoduli are stabilized at the origin, and there is metastable DSB.  Another modification -- which is potentially more suitable for model building -- is to deform the theory by $W_{tree}\supset m\Tr\,\Phi_{00}^2$ \GKKS. Since this is a nonrenormalizable interaction in the UV, $m$ is naturally small; as shown in \GKKS, this makes it possible to balance the two-loop runaway potential against this tree-level stabilizing potential and obtain a meta-stable vacuum at $\Phi_{00}\ne 0$ where the R-symmetry is completely broken.

\subsec{SQCD with weakly gauged flavor symmetry -- an example with gauge messengers}

We start with the SQCD theory considered in \ISS, and reviewed in sect. 6.1, where all $N_f$ flavors are give the same mass $m=m_Q$.  There is a global $SU(N_f)\times U(1)_B\cong U(N_f) $ symmetry,
and we here consider gauging some subgroup $G\subseteq SU(N_f)$ of the flavor symmetry. We will very weakly gauge this subgroup, $g\ll 1$,   so that $\Lambda _{SU(N_c)}\gg \Lambda _{G}$.  For energies below $\Lambda _{SU(N_c)}$, we dualize $SU(N_c)\to SU(N=N_f-N_c)$ and, as in \ISS, there are metastable SUSY-breaking vacua, given at tree level by  
\phivarphipm.

Let us first consider the case that $G=SU(N_f)$. (In the next subsection we will analyze the case of proper subgroups.) Some preliminary analysis of this theory appeared in \refs{\ISS, \ForsteZC}.  The classical vacua are still given by \phivarphipm\ for $g\neq 0$, though the
$SU(N_f)$ D-terms lift some of the $g=0$ pseudomoduli at tree level.  The 
SUSY-breaking F-terms are still given by \Fvevs.  Our interest here will be in the theory with $k$ added $SU(N_f)$ flavors, $\rho^i\in(1, N_f)$, $\widetilde\rho_i\in(1,\overline {N_f})$, $i=1,\dots,k$, with $k$ in the range $N_f+N_c<k<3N_f-N_c$, so that the $SU(N_f)$ gauge coupling is asymptotically free in the UV $SU(N_c)\times SU(N_f)$ theory, and IR free in the $SU(N)\times SU(N_f)$ low-energy dual \ForsteZC.  The dual theory has 
\eqn\wsqcdflavor{W=h\Tr \Phi \varphi \widetilde \varphi +m_Q\Lambda \Tr \Phi +m_\rho \rho \widetilde \rho.}  
We are interested in the case where $m_\rho =0$.   The added $SU(N_f)$ matter fields in this case 
lead to additional 
pseudomoduli, given by the expectation values of $\rho $ and $\widetilde \rho$ along the tree-level D-flat directions, Higgsing $SU(N_f)$ (or $U(N_f)$). The effective potential for these pseudomoduli had not yet been computed in the literature.  

We here highlight a key point about this theory: because $\hat\Phi$ in \phivarphipm\ is charged under $SU(N_f)$, the non-zero $F$-term for this charged field implies that the model has gauge messengers, as discussed in section 5.4. Thus the $\rho$, $\tilde\rho$ pseudo-moduli are lifted at one instead of two loops. Indeed, a direct computation of the one-loop Coleman-Weinberg potential exhibits the dependence on the $\rho$ and $\widetilde \rho$ expectation values.  Alternatively, we can use apply the discussion in section 5.4 to see that the one-loop potential is non-vanishing at large $\rho$, $\tilde\rho$.

Consider the pseudomodulus direction where  $\rho$ and $\widetilde \rho$ each have a single large entry, $\rho _1$. This Higgses $SU(N_f)$ to $SU(N_f-1)$ at the threshold scale $g\rho _1$. Under $SU(N_f-1)$, $\Phi$ decomposes into an adjoint $\phi_A$, $A=1,\dots,{\rm dim}(SU(N_f-1))$; two singlets $\phi_{0'}$ and $\phi _0$; and a fundamental plus anti-fundamental. The latter two do not participate in the SUSY breaking so we will set them to zero henceforth. The decomposition of $\Phi$ into the remaining fields is:
\eqn\Phidecompose{
 \Phi = \phi_A T^A + \phi_{0'} T^{0'} + \phi _0 N_f^{-1/2} \unit_{N_f} 
 }
where $T^A$ are the generators of $SU(N_f-1)$ (as $N_f\times N_f$ matrices), and $T^{0'}$ is proportional to the $SU(N_f-1)$ (but not the $SU(N_f)$) identity. These generators are all normalized to have $\Tr\,(T^A)^2 = \Tr\,(T^{0'})^2 = 1$ so that $\phi_A$, $\phi_{0'}$ and $\phi _0$ are canonically normalized fields. The corresponding F-terms are then given by decomposing \Fvevs\ as 
\eqn\FPhidecompose{  F_{\Phi }=F_{\phi _A}T^A+F_{\phi_0'}T^0 +F_{\phi _0} N_f^{-1/2}\unit_{N_f}.
}
In general these F-terms will all be non-zero.   Then according to \vgmhiggs, the effective potential for $\rho_1$ in the regime \yrange\ is
\eqn\vgmhiggseg{
V_{eff} \approx -2\Big(|F_{\phi_A}|^2\Delta c(\phi_A) + |F_{\phi_{0'}}|^2\Delta c(\phi_{0'})\Big)\kappa_g \log {|\rho_1|\over m_0}
}
Notice that $F_{\phi _0}$ does not contribute since it is neutral under $SU(N_f)$ and $SU(N_f-1)$. The change in the quadratic Casimir invariants is simply
\eqn\deltacgmh{\eqalign{
 & \Delta c(\phi_A) = c(SU(N_f)\,\, {\rm adj}) - c(SU(N_f-1)\,\, {\rm adj}) = 1 \cr
 & \Delta c(\phi_{0'}) = c(SU(N_f)\,\, {\rm adj}) - c(SU(N_f-1)\,\, {\rm sing}) = N_f. \cr
}}
Substituting back into \vgmhiggseg, we conclude that $V_{eff}$ is indeed a downward-sloping function of the pseudomodulus field $\rho_1$, indicating ``bad" runaway behavior. 

Let us make two comments on this result. First, we have been intentionally vague about the precise forms of $F_{\phi_0}$ and $F_{\phi_A}$, since these will depend on how the $\rho$ vev is aligned with the SUSY-breaking pattern \phivarphipm-\Fvevs. However, we see from this analysis that the conclusion of runaway behavior is robust and does not depend on the detailed form of the F-term vevs.
Second, let us mention that the coefficient in \vgmhiggseg, and in particular its sign, can also be obtained from the fact that the coefficient of $\ln m_0$ is the same as the coefficient of $\ln M_{cutoff}$ in the Coleman-Weinberg potential: ${\partial V\over \partial \ln m_0}={\partial V_{CW}^{(1)}\over \partial \ln M_{cutoff}}=-{1\over 32\pi ^2}{\rm Str}\CM ^4$.  Indeed the result \vgmhiggseg\ is reproduced upon evaluating the leading $\CO (|F|^2)$ contribution to ${\rm Str}\CM ^4$ in the large $|\rho _1|$ limit, using the appropriate gauge messenger spectrum (which can be found as in \GiudiceNI).

\subsec{SQCD with weakly gauged flavor symmetry -- no gauge messengers}

Let us now consider the same model as the previous subsection except that
instead of gauging the entire $SU(N_f)$ flavor symmetry, we gauge an $SU(K)$ subgroup which is sufficiently small, and aligned, such that such that $F_{\Phi }$ in \Fvevs\ is gauge neutral.  This of course leads to qualitatively different behavior, since now there are no longer any gauge messengers present to lift the pseudomoduli at one loop.

There are now two qualitatively different possibilities for how the $SU(K)$ gauge group is aligned.  The 
expectation values \phivarphipm\ break $SU(N)\times SU(N_f)\to SU(N)_D\times SU(N_f-N)$, and the two possibilities are that the $SU(K)$ can align inside either $SU(N)_D$ or inside $SU(N_f-N)$ (assuming that $K\leq N$, or $K\leq N_f-N$, respectively).  The qualitative difference is because \Fvevs\ leads to tree-level SUSY-split masses for only the last $N_f-N$ flavors.  So if $SU(K)$ aligns with the first $N$ entries in \Fvevs, then the messenger fields are neutral under $SU(K)$, whereas if the $SU(K)$ aligns with the last $N_f-N$ entries in \Fvevs\ the messengers $\varphi$ are charged under $SU(K)$.  In either case, we are here considering the case where the F-terms \Fvevs\ are $SU(K)$ gauge singlets, so that there are no gauge messengers.

Let us first consider the case where $SU(K)$ aligns within the $SU(N_f-N)$, i.e.\ the last $N_f-N$ flavors in \Fvevs.  Since there are then messengers $\varphi$ with SUSY-split masses, and charged under $SU(K)$, the $\rho$ pseudomoduli in this case couple as the Higgsing pseudomoduli of section 5.2.  As discussed there, such pseudomoduli are lifted at two loops, and such pseudomoduli are good -- their potential safely stabilizes them.  Indeed, consider the range \yrange, where their
potential can be read off from \vtwoex.  Consider again the pseudo-D-flat direction where $\rho$ and $\widetilde \rho$ each have a single large entry, $\rho _1$, Higgsing $SU(K)$ to $SU(K-1)$.   

We then have from \twoloopifgii\
\eqn\ctwoexgfli{\Delta \Omega _X^{(2)}=-8\kappa _{h_X}\kappa _gT(r_\varphi)(|G|-|G'|)=-4{\kappa _h\over N_f-N}\kappa _g  N(2K-1),}
where $h$ is as in \wsqcd\ and $h_X=h/\sqrt{N_f-N}$ comes from writing the F-terms \Fvevs\ in terms of a canonically normalized $SU(N_f-N)$ singlet field $X$.  As in \vtwoex, the potential is then 
\eqn\vtwoexgfl{V_{eff}(\rho _1)\approx |f|^2 \kappa _{h} \kappa _gN(2K-1)\ln ^2{|g\rho _1|^2\over m_0^2},}
which is an increasing function of $|\rho _1|$; the potential safely stabilizes these pseudomoduli. 
Allowing for several flavors of $\rho$ and $\widetilde\rho$ with widely separated large expectation values yields a sum of terms like \vtwoexgfl, safely lifting all of these Higgsing  pseudomoduli.

We now briefly summarize the  case where $SU(K)$ instead aligns inside the unbroken $SU(N)_D$, i.e.\ within the first $N$ entries in \Fvevs.   The SUSY-split messenger components of $\varphi$ are now $SU(K)$ neutral, and the $SU(K)$ charged components of $\varphi$ have SUSY masses.   These components interact via the superpotential terms coming from \wsqcd, with coupling $h$. The upshot is that the $\rho$ pseudomoduli in this case are 
first lifted at {\it three} loops.    The three-loop effective potential in the range \yrange\ is approximated by \Vyapproxii\ with $\Delta \Omega _X^{(3)}={\partial \gamma _X\over \partial h_X}{\partial \beta_{h_X}\over \partial h}\Delta \beta _h<0$, where the sign comes from the $SU(K)$ gauge contribution to $\Delta \beta _h$. It then follows from \summary\ that the $\rho$ pseudomoduli in this case are bad.

To summarize, if the gauged flavor group is the entire $SU(N_f)$, or more generally an $SU(K)$ subgroup which is not aligned within either the first $N$ entries or the remaining $N_f-N$ entries in \Fvevs, then there are gauge messengers, and the $\rho$ pseudomoduli are lifted at one loop and are bad, with a perturbative runaway.  
If the gauged $SU(K)$ flavor subgroup is entirely within the last $N_f-N$ entries in \Fvevs, then the $\rho$ pseudomoduli are lifted at two loops and are good.  Finally, if the gauged $SU(K)$ is entirely within the first $N$ entries in \Fvevs, then the $\rho$ pseudomoduli are lifted at three loops and are bad.

\subsec{Example with a saxion-type pseudomodulus: $SU(N_c)$ with symmetric tensor and antifundamentals}

We now take the UV theory to be an $SU(N_c)$ gauge theory with symmetric tensor $S$ and $N_f=N_c+4$ antifundamentals $\widetilde Q_i$, and we attempt to break supersymmetry by turning on the tree-level superpotential $W_{tree}=\Tr \lambda ^{ij}  S\widetilde Q_i\widetilde Q_j$.  For simplicity, we take $\lambda ^{ij}=\lambda \delta ^{ij}$, preserving an $SO(N_f)\subset SU(N_f)$ flavor symmetry, along with a $U(1)_R$ symmetry with $R(S)=-2+{4\over N_c}$ and $R(\widetilde Q)=2-{2\over N_c}$. 
This theory was originally considered long ago \AffleckXZ, where it was noted to have an interesting pseudo-flat direction, labeled by $\ev{\det S}$, along which $\ev{\widetilde Q}=0$ and $\ev{S}=a\unit _{N_c}$.  Far from the origin in this direction, $\ev{\det S}\gg \Lambda ^{N_c}$,  there is a non-perturbative runaway superpotential which pushes $\det S\to \infty$ \AffleckXZ:
\eqn\wdynsymten{W_{dyn}=c\left({(\Lambda ^{2N-3})^2 \det \lambda \over \det S}\right)^{1/(N_c-2)}}
(To obtain \wdynsymten, note that $\ev{S}=a\unit _{N_c}$ Higgses $SU(N_c)$ to $SO(N_c)$, and gives the $\widetilde Q$ mass $m_{\widetilde Q}=\lambda a$.  Then \wdynsymten\ is generated by gaugino condensation in the low-energy $SO(N_c)$ Yang-Mills theory.) It was speculated in \AffleckXZ\ that there might be a metastable minimum at smaller values of  $\det S$, perhaps either in the $\ev{S}\gg \Lambda$ regime, or for $S$ nearer the origin.  

The theory near the origin can now be analyzed using its known magnetic dual \PouliotSK: an $SO(8)$ gauge theory with $N_f$ matter fields $\varphi \in {\bf 8}_v$, one matter field $p\in {\bf 8}_s$, $\half N_f(N_f+1)$ singlets $\Phi$ (with $\Phi =\Phi ^T$), and one more singlet $Z$,  
with superpotential 
\eqn\wpsdual{W_{dual}=h(\Tr \Phi \varphi \varphi -\mu ^2 \Tr \Phi + Z p^2),}
where, as before, we take the couplings to be the same for simplicity.  In terms of the UV theory, $\Phi = \Lambda ^{-2}S\widetilde Q\widetilde Q$, $Z=\Lambda ^{1-N_c}\det S$, and $f=-h\mu ^2=\lambda \Lambda ^2$.  The $SO(8)$ magnetic dual theory  is IR free for  $N_f\geq 17$.  The small parameter \fepsilon\ is $\epsilon =\lambda$, so we need to take $|\lambda|\ll 1$.    The first two terms in \wpsdual\ lead to $F_{\Phi}\neq 0$ via the rank-condition supersymmetry breaking \ISS, with tree-level vacuum $V_0=(N_f-8)|h\mu ^2|^2\sim (N_f-8)|\lambda \Lambda ^2|^2$.  Indeed, this sector of the theory is the IR dual of an $SO(N_c)$ gauge theory with $N_f=N_c+4$ massive fundamentals which, along with the field $Z$, is the low-energy electric theory obtained for large $Z$.    Ignoring the $Z$ pseudomodulus, this low-energy theory would have metastable DSB vacua at $\ev{\Phi }=0$, with $\ev{\varphi}\neq 0$ (breaking $SO(8)_{gauge}\times SO(N_f)_{flavor}\to SO(8)_D\times SO(N_f-8)$), just as in \ISS.  

However, the additional $Z$ pseudomodulus of the low-energy theory \wpsdual\ spoils the metastable DSB minimum in the other pseudomoduli.     The field $Z$ is the same interesting pseudomodulus,  with the non-perturbative runaway for $|Z|\gg |\Lambda|$, found in \AffleckXZ.  For $|Z|\ll |\Lambda|$, the IR free magnetic dual $SO(8)$ theory reveals a perturbative runaway,  as the field $Z$ is of the ``saxion" type, like $\Phi _3$ in \wsaxion.  As we have argued, such pseudomoduli develop a perturbative runaway potential at three loops, and eventually $Z$ slides to the UV cutoff of the low-energy theory, $Z\sim \Lambda$, where all bets are off. In the regime \yrange\ of  $|\mu|\ll |X|\ll |\Lambda|$, the effective potential is given by \vthreeex\ (where $g$ is the dual $SO(8)$ gauge coupling):
 \eqn\vthreeexis{V_{eff}(X)\approx -{56\over 3}V_0\Kappa\kappa_g^2\left(\ln {|Z|^2\over |\mu ^2|}\right)^3}
 since $b=17-N_f$ above the $Z$ threshold, and below $p$ gets a mass, so $b'=18-N_f$. 

Because of this perturbative runaway, we expect that this theory does {\it not} have a metastable dynamical supersymmetry breaking near the origin\foot{We have not computed the three-loop $V_{eff}^{(3)}$ in the range $|X|\sim |\mu|$, and in principle there could be a metastable minimum in this range very close to the origin.  Even if that were the case, such a hypothetical minimum would likely not be sufficiently long-lived to be viable.}.   All evidence points toward this theory having everywhere the runaway to $\ev{S}\to \infty$, starting with the perturbative magnetic runaway for $\ev{S}\ll \Lambda$, and ending with the nonperturbative electric potential from \wdynsymten\ for $\ev{S}\gg \Lambda$, rather than any metastable DSB vacuum. 

\subsec{Modifications of the above example, which do have metastable DSB near the origin}

We can still modify the electric $SU(N_c)$ theory to remove the runaway by hand, and obtain a model of DSB.    We add to the electric theory a gauge singlet field $\Sigma$, and take 
 \eqn\wtreeelec{W_{tree}=\lambda \Tr S\widetilde Q\widetilde Q+{c\over M_p^{N_c-2}}\Sigma\ \det S,}
 where $M_p$ is the scale of some UV completion or other dynamics (suppose $M_p\gg |\Lambda|$) and $c$ is a dimensionless constant\foot{Replacing the last term in \wtreeelec\ with $c\ (\det S)^2/M_p^{2N_c-3}$ is qualitatively similar to the case described above.  An alternative is to replace the last term in  \wtreeelec\ with $c\ \det S/M_p^{N_c-3}$, which also halts the $\ev{Z}\to \infty$ runaway.  
 This theory does admit metastable DSB vacua,  related to those of $SO(N_c)$ with $N_f=N_c+4$, but the necessary condition \fepsilon\ becomes the requirement that $\lambda$ be unnaturally small: $\lambda \ll (\Lambda /M_p)^n$ where $n>0$ depends on $N_c$ ($n\to 2$ for large $N_c$).}. In the IR free magnetic dual, the superpotential is 
 \eqn\wpsdual{W_{dual}=h(\Tr \Phi \varphi \varphi -\mu ^2\Tr \Phi +Z p^2)+m_Z\Sigma Z,}
 where $m_Z=c\Lambda (\Lambda/M_p)^{N_c-2}$.   As before, we take $N_f>17$, so the magnetic theory is IR free.  The first two terms then lead to rank-condition supersymmetry breaking, with $V_0=(N_f-8)|h\mu ^2|^2$.  The last term gives $Z$ a tree-level supersymmetric mass $m_Z$, so $Z$
 is no longer a pseudomodulus; the runaway direction of the previous subsection has been eliminated.  The new field $\Sigma$ leads to a new pseudomodulus, but this one is dynamically stabilized.  Indeed, integrating out the massive field $Z$, its equation of motion sets $\Sigma \sim p^2$, so the new pseudomodulus is of the Higgsing type; it is a pseudo-D-flat direction along which the spinor $\ev{p}$ gets an expectation value, Higgsing $SO(8)$ to $SO(7)$.  This pseudo-D-flat direction is lifted at two loops, with $V_{eff}^{(2)}$ minimized at the origin, $\ev{\Sigma}=\ev{p}=0$.  Near the origin, $V_{eff}\supset m_p^2p^\dagger p \sim m_p^2 \sqrt{\Sigma ^\dagger \Sigma}$, with $m_p^2>0$. 
In the range \yrange\ of the pseudomodulus $Y\sim g_{SO(8)}\sqrt{c\Lambda \Sigma}$, the effective potential is given by \vtwoex\
 \eqn\vtwoexex{V_{eff}(\Sigma)\approx 16V_0\Kappa\kappa_g\Delta c(r_\varphi)\left(\ln {|Y|^2 \over |\mu|^2}\right)^2,}
with $\Delta c(r_\varphi)=({7\over 2}-{21\over 8})$.  So in this limit too the potential is an increasing function of $|\Sigma|$.  

A different modification of the example of the previous subsection is to weakly gauge the $SO(N_f)$ symmetry, with gauge coupling $g'$.  The $\Phi$ of the dual theory \wpsdual\ decompose into an $SO(N_f)$ adjoint and singlet, $\Phi = \phi _AT^A+\phi _0 N_f^{-1/2}\unit _{N_f}$, as does $F_\Phi$, which is given as in \Fvevs\ (with $N=8$).  Since $F_{\phi _A}\neq 0$, there are gauge messengers, which
lifts the  saxion $Z$ at two loops.  For $Z$ in the range \yrange, its potential is given as in \gmesssaxv
\eqn\gmessaxvis{V_{eff}(Z)\approx 32{(N_f-8)(N_f-2)\over N_f}|f|^2\kappa _{g'}^2\left(\log  {|Z|\over |\mu |}\right)^2,}
where we used  $\sum _A |F_{\phi _A}|^2 =8(N_f-8)|f|^2/N_f$.  This potential safely stabilizes the saxion, so the theory of the previous subsection can have viable DSB upon gauging the $SO(N_f)$ flavor symmetry. (The $SO(N_f)$ can run to strong coupling in the IR, unless additional $SO(N_f)$ charged matter is added.  Such matter can lead to bad Higgsing pseudomoduli, as in section (6.3).)

\subsec{Examples with incalculable pseudomoduli potentials: Kutasov-type dualities}

As in \refs{\KutasovVE, \KutasovNP, \KutasovSS} there are many duality examples based on matter fields in multiple representations, with an added tree-level superpotential for some of the representations; see e.g. \IntriligatorAX\ for additional examples.  In all of the duals, some  moduli enter only via power-law irrelevant terms.  If supersymmetry is broken, these become irrelevantly coupled 
 pseudomoduli, of the type $\Phi _4$ in \wfour.  Thus {\it none} of these examples can have calculable metastable DSB -- they are all inconclusive.   

Consider, for example,  the original example of \KutasovVE.  The electric theory is $SU(N_c)$ SQCD, with $N_f$ fundamental flavors and an added adjoint $X$,  with $W_{tree}=\Tr X^3+\lambda \Tr QX\widetilde Q$.  The term with coupling $\lambda$ has been added  to try to dynamically break supersymmetry.  The dual theory \KutasovVE\ has gauge group $SU(N=2N_f-N_c)$,  with adjoint $Y$,
$N_f$ fundamental flavors $\varphi$ and $\widetilde \varphi $, and gauge singlets $\Phi _0=Q\widetilde Q/\Lambda$ and $\Phi _1=QX\widetilde Q/\Lambda^2$, with
\eqn\wkdual{W_{dual}=hTr \Phi _1 \varphi \widetilde \varphi+f\Tr \Phi _1+\Tr Y^3+{a\over \Lambda} \Tr \Phi _0\varphi Y\widetilde \varphi,}
where the dimensionless couplings $h$ and $a$ are $\CO (1)$ at the cutoff and $f=\lambda \Lambda ^2$.  The theory is IR free for $N_f<2N_c/3$ and the small parameter $\epsilon$ in \fepsilon\ is here given by $\epsilon =\lambda$.

The first two terms in \wkdual\ give the rank-condition supersymmetry breaking sector \ISS, with $F_{\Phi _1}\neq 0$, and $\ev{\varphi}=\ev{\widetilde \varphi}^T\neq 0$.  The mass spectrum and pseudomoduli potential for the components of $\Phi _1$ and $\varphi$ are identical to that of the SQCD example of \ISS. The additional fields $Y$ in \wkdual, and also $N^2$ components of the $\Phi _0$ fields, get calculable tree-level supersymmetric masses,  $m^2_{calc}\sim |a \epsilon \Lambda /h|^2\sim |a^2 \epsilon ^2\ln |\epsilon| \Lambda ^2|$ from the $\widetilde \varphi\varphi \neq 0$.   These calculable tree-level masses just barely robust against the unknown $m^2_{incalculable}\sim |\epsilon \Lambda |^2$, thanks to the $|h|^{-2}\sim - \ln |\epsilon|$ enhancement.  

But there are remaining pseudomoduli components of $\Phi _0$,  which can be first lifted at one loop.  Because
they  enter into the low-energy theory only via the power-law irrelevant last term in \wkdual, it follows from our general discussion in section 3 that the effective potential for these pseudomoduli is incalculable, $m_{calc}^2\leq m^2_{incalculable}$.  For example, the one-loop calculable contribution  $m^2_{calc}\sim |a\epsilon \Lambda |^2$ is of the same order as the incalculable contributions, from terms like $K_{eff}\supset c|\Lambda |^{-2}\Phi _1^\dagger \Phi _1 \Phi _0^\dagger \Phi _0$.  So we cannot determine whether the $\Phi _0$ pseudomoduli are stabilized in the region $|\Phi _0|<|\Lambda|$, or if they instead develop a dangerous runaway to larger values of $\Phi _0$, where the low-energy analysis is inapplicable.  It is thus inconclusive whether or not this theory dynamically breaks supersymmetry.   

One can still modify the UV theory to eliminate, by hand, the dangerous pseudomoduli by adding mass terms for them.  In the examples at hand, this fix has already  been implemented in the literature, with $\Phi _0$ given mass via  superpotential term $\Tr \Sigma Q \widetilde Q \to \Tr \Sigma \Phi _0$ (with added gauge singlets $\Sigma$) \OoguriPJ\ or alternatively $Tr (Q\widetilde Q)^2\to \Tr \Phi _0^2$  \AmaritiVK.

\subsec{Analogs of SQCD with $N_f=N_c+1$: IR free theories without gauge fields}

 While there is no general classification of which supersymmetric gauge theories have IR free low-energy duals (as opposed to an interacting SCFT), some classes of theories have been well mapped out. For example, there is a classification of the ``s-confining" ${\cal N}=1$ theories with simple gauge group and $W_{tree}=0$.  These are the theories analogous to SQCD with $N_f=N_c+1$ \SeibergBZ: the low energy IR free fields have only superpotential, and no (dual) gauge interactions.  Another example is 
$Sp(N_c)$ with $N_f=N_c+2$ flavors \IntriligatorNE.   Many analogous theories were summarized in 
\CsakiZB.    The basic fields of the 
IR free theory are  all gauge invariant composites of the UV matter, ignoring the classical relations,
and there is a $W_{dyn}$ whose F-term equations give the classical relations.  
Adding a linear term in one of the IR free fields  can potentially break supersymmetry.  

As shown in \ISS, the $SU(N_c)$ theory with $N_f=N_c+1$ has {\it calculable} metastable dynamical supersymmetry breaking, whereas the $Sp(N_c)$ theory with $N_f=N_c+2$ does not.  The difference is 
that all pseudomoduli of the $SU(N_c)$ theory enter into cubic terms in $W_{dyn}$, whereas 
pseudomoduli of the $Sp$ theory  couple via power-law irrelevant terms in $W$, so their effective potential is incalculable.   Again, if there are {\it any} incalculable pseudomoduli, one cannot determine whether or not the theory has DSB -- it depends on the sign of the incalculable higher-order Kahler potential terms.

A scan of the other examples in  \CsakiZB\ reveals that the $SU(N_c)$ SQCD is a rather special example.  The other examples more generically have many fields which appear in $W_{dyn}$ via terms which are power-law irrelevant, which will become incalculable pseudomoduli if a sector of the theory 
breaks supersymmetry.   

As an example, consider $SU(N_c)$ with one flavor of antisymmetric tensor, $A$ and $\widetilde A$, and $N_f=3$ fundamental flavors, $Q$ and $\widetilde Q$.  The IR free theory is discussed in \CsakiZB.  
If we add $W_{tree}=m_AA\widetilde A+m_Q\Tr Q\widetilde Q$, the $m_Q$ term leads to a rank-condition supersymmetry breaking sector (with $\Phi _1= Q\widetilde Q/\Lambda$ and $\varphi = \widetilde A(A\widetilde A)^{\half N_c-1}Q^2$), so it is possible that this theory has a metastable DSB vacuum near the origin.    But many pseudomoduli, e.g. $T_1=A\widetilde A$, couple only via superpotential terms of quartic and higher order.   Thus they are not reliably stabilized within the low-energy effective theory, and the DSB vacuum requires an assumption about the sign of non-calculable terms in the Kahler potential.  Such potentially dangerous pseudomoduli can still can be stabilized by hand, by modifying the UV theory to give them masses, to obtain a theory with (metastable) DSB vacua.

The scan of these classes suggest that calculability is perhaps not generic. 

\bigskip

\noindent {\bf Acknowledgments:}

We especially thank Nati Seiberg for early collaboration on this project, and for many interesting discussions and useful comments.  We would also like to thank Nima Arkani-Hamed for helpful discussions. The research of KI  and MS is supported in part by UCSD
grant DOE-FG03-97ER40546.  The research of DS is supported
in part by DOE grant DE-FG02-90ER40542.  KI would like to thank the IAS, the IHES, the ENS Paris, the Newton Institute, and the International Centre for Theoretical Studies of the TIFR Mumbai for hospitality and support over the times where parts of this work were completed.  

\appendix{A}{Deriving the Leading-Log Effective Potential}

In this appendix, we derive the formula \Vyapproxii\ for the leading-log effective potential in the large field regime \yrange.  

The statement that we can approximate $\Omega_X$ with leading logs is the statement that we have a good power-series expansion of the form
\eqn\Zleadlog{
\Omega_X(t) \approx C_0 + C_1 \kappa (t-t_\Lambda) + {1\over2!}C_2(\kappa (t-t_\Lambda))^2 + \dots
}
with $\kappa\ll \kappa(t-t_\Lambda) \ll 1$ and $t=\log Q/m_0$. Note that we are expanding $\Omega_X$ around the UV scale $\Lambda$ but we are defining the RG time with respect to the IR scale $m_0$; the reasons for this will be apparent in a moment.  Because the effect we are after is, in fact, a leading log effect, we drop terms of $\CO(\kappa ^n(t-t_\Lambda)^m)$ with $m<n$ and (since there are not terms with $m>n$) we only keep the terms with $n=m$.

Now suppose that at $t=t_\Phi$, the $n$th derivative of $\Omega_X$ is discontinuous with all lower-order derivatives still continuous. Then for $t< t_\Phi$, the expansion \Zleadlog\ becomes
\eqn\Zleadlogii{
 \Omega_X(t) = C_0' + C_1' \kappa (t-t_\Lambda) + {1\over2!}C_2'(\kappa (t-t_\Lambda))^2 + \dots \qquad\qquad (t<t_\Phi)
 }
with the new coefficients $C_i'$ satisfying
\eqn\Zdiscii{
 \sum_{k=i}^\infty {1\over (k-i)!} C_k (\kappa (t_\Phi-t_\Lambda))^{k-i} =  \sum_{k=i}^\infty {1\over (k-i)!} C_k' (\kappa (t_\Phi-t_\Lambda))^{k-i} ,\qquad i=0,\dots,n-1.
}
The Taylor expansion coefficients $C_i$ of $\Omega_X$ around the UV scale are independent of $t_\Phi$. Then \Zdiscii\ yields a system of equations which determine the $t_\Phi$ dependence of the IR Taylor coefficients $C_0',\dots,C_{n-1}'$.  It is straightforward to check that \Zdiscii\ are solved by
\eqn\Zdiscsol{
C_i' = C_i - {1\over (n-i)!}(C_n-C_n')(-\kappa (t_\Phi-t_\Lambda))^{n-i} +\dots\qquad (i=0,\dots,n-1)
}
where $\dots$ denote terms that are higher order in $\kappa$. Plugging this into \Zleadlogii, we see that 
\eqn\Zleadlogiii{\eqalign{
\Omega_X(0) 
 &= const. - \sum_{i=0}^n {1\over i!(n-i)!}\kappa^n(-1)^{n}(C_n-C_n')(t_\Phi-t_\Lambda)^{n-i}t_\Lambda^i+\dots\cr
  &= const. - {1\over n!}(C_n-C_n')(-\kappa t_\Phi)^n+\dots.
}}
This reproduces \ZXupshot\ after we identify $\Delta\Omega_X^{(n)}\equiv{d^n\Omega_X\over dt^n}\big|^{t_\Phi^+}_{t_\Phi^-}=\kappa^n(C_n-C_n').$

\listrefs
\end